\documentclass[final]{siamltex}
\usepackage{psfrag}
\usepackage{epsfig}
\usepackage{subfigure}

\newcommand{\comments}[1]{}
\newcommand{\ba}{\begin{eqnarray}}
\newcommand{\ea}{\end{eqnarray}}
\newcommand{\be}{\begin{equation}}
\newcommand{\ee}{\end{equation}}

\title{Numerical Implementation of Streaming Down the Gradient: Application to Fluid Modeling of Cosmic 
Rays and Saturated Conduction}

\author{Prateek Sharma \thanks{Chandra Fellow, Theoretical Astrophysics Center and Astronomy Department, University of California, Berkeley, CA 94720 ({\tt psharma@astro.berkeley.edu }).}
        \and Phillip Colella and Daniel F. Martin \thanks{Applied Numerical Algorithms Group, Lawrence Berkeley National Laboratory, Berkeley, CA 94720}}

\begin{document}

\maketitle

\begin{abstract}

The equation governing the streaming of a quantity down its gradient superficially looks similar to the simple constant velocity
advection equation. In fact, it is the same as an advection equation if there are no local extrema
in the computational domain or at the boundary. However, in general when there are
local extrema in the computational domain it is a non-trivial nonlinear equation. 
The standard upwind time evolution with a 
CFL-limited time step results in spurious oscillations at the grid scale. These oscillations, which originate
at the extrema, propagate throughout the computational domain and are undamped even at late times. 
These oscillations arise because of unphysically large fluxes leaving (entering) the maxima (minima) with the standard CFL-limited
explicit methods. Regularization
of the equation shows that it is diffusive at the extrema; because of this, an explicit method for the regularized equation with $\Delta t \propto \Delta x^2$
behaves fine. We show that the implicit methods show stable and converging results with $\Delta t \propto \Delta x$; however, surprisingly, even implicit 
methods are not stable with large enough timesteps. In addition to these subtleties in the numerical implementation, the solutions to the
streaming equation are quite novel: non-differentiable solutions emerge from initially smooth profiles; the solutions show transport over large
length scales, e.g., in form of tails. The fluid model for cosmic rays interacting with a thermal plasma 
(valid at space scales much larger than 
the cosmic ray Larmor radius) is similar to the equation for streaming of a quantity
down its gradient, so our method will find applications in fluid modeling of cosmic rays.

\end{abstract}


\pagestyle{myheadings}
\thispagestyle{plain}
\markboth{SHARMA, COLELLA, and MARTIN}{Numerical Implementation of Streaming Down the Gradient: Application to Fluid Modeling of Cosmic Rays}

\section{Introduction}
Cosmic rays (energetic particles moving close to the speed of light) are an important dynamic and energetic component of our Galaxy.
The average cosmic ray pressure in our Galaxy is comparable to the kinetic, thermal, and magnetic pressures \cite{Boulares1990}.
It is important to study the interaction of cosmic rays with the thermal plasma to understand the dynamics and the energy balance in our 
Galaxy and in other astrophysical objects, e.g., clusters of galaxies.

Cosmic rays are collisionless, with the Coulomb collision time $\gg$ the dynamical timescales. However, cosmic rays do not just leave the 
system at the speed of light. Cosmic rays are effectively coupled to the thermal plasma by self-generated Alfv\'en waves that scatter them (e.g., \cite{Kulsrud2005}; efficient scattering occurs only when the cosmic rays have a sufficiently large number density; this is usually the case 
for the majority of cosmic ray particles in the Galaxy; e.g., \cite{Wentzel1971}). 
These waves are generated by a fast, self-generated instability that arises when the cosmic ray streaming velocity 
(velocity of the cosmic ray fluid relative to the thermal plasma) exceeds the local Alfv\'en speed. As a result of this instability, the cosmic ray 
fluid streams at the local Alfv\'en speed (down the cosmic ray pressure gradient) with respect to the thermal plasma. Other mechanisms, like
MHD turbulence (e.g., \cite{Yan2004}) and non-resonant instabilities (e.g., \cite{Bell2004}) can also scatter cosmic rays but the qualitative picture 
of streaming relative to the thermal plasma should hold even in these cases. For simplicity, we focus on the self-generated streaming instability and 
assume that the cosmic ray fluid streams relative to the thermal plasma at the local Alfv\'en speed.

Although a fully kinetic description is necessary to study
the excitation of Alfv\'en waves (at the cosmic ray Larmor radius scale) and their damping on the thermal plasma, a fluid description,  where the
cosmic ray fluid streams at the local Alfv\'en speed relative to the thermal plasma, suffices at scales much larger than the cosmic ray Larmor radius 
(e.g., the typical cosmic ray Larmor radius is $\ll$ the thickness of our Galaxy, and hence a fluid model of cosmic rays is sufficient to study 
the large scale dynamics and energetics).
The equation governing the cosmic ray pressure is (generalization of 1-D equations from \cite{McKenzie1982})
\be
\label{eq:pcr}
\frac{\partial p_c}{\partial t} + \vec{\nabla} \cdot \left (p_c \vec{u}\right ) + 
\vec{\nabla} \cdot ( \frac{4}{3} p_c \vec{v}_s)  = -p_c \frac{\vec{\nabla} \cdot \vec{u}}{3} - 
\frac{| \vec{v}_s \cdot \vec{\nabla} p_c|}{3},
\ee
where $p_c$ is the cosmic ray pressure and $\vec{u}$ is the thermal plasma velocity. The streaming velocity $\vec{v}_s= 
-{\rm sgn}(\vec{B} \cdot \vec{\nabla}p_c) \vec{B}/\sqrt{4\pi \rho}$,  where $\vec{B}$ is the magnetic field vector 
and $\rho$ is the plasma density, is equal to the Alfv\'en velocity, but down the cosmic ray pressure gradient. 
Equation (\ref{eq:pcr}) can be derived by replacing $\vec{u}$ by $\vec{u}+\vec{v}_s$ in the standard internal
energy equation for a $\gamma=4/3$ (relativistic) fluid. The last term on the right hand side of Eq. (\ref{eq:pcr})
is a loss term since cosmic rays work in pushing the thermal plasma as they stream down their own
pressure gradient; there is a compensating heating term in the plasma internal energy equation. This irreversible heating 
can be understood in the kinetic framework: cosmic ray fluid streaming at speeds slightly larger than the local Alfv\'en speed
results in (usually small amplitude) Alfv\'en waves at the cosmic ray Larmor radius scale. These waves efficiently scatter the 
cosmic ray particles. In addition to elastic scattering,
cosmic rays also lose their energy to Alfv\'en wave packets because the self-generated waves are moving, on average (in the fluid sense), 
away from the cosmic ray particles (this process is the inverse of Fermi acceleration). Thus, a part of cosmic ray energy is converted into 
Alfv\'en waves which, in a steady state, damp on the thermal plasma. In fluid modeling (valid at large time and space scales), it is
assumed that the cosmic rays directly lose their energy to the thermal plasma via the last term in Eq. (\ref{eq:pcr}).

Eq. (\ref{eq:pcr}) can be solved by operator splitting, where each
term is updated individually. 
The terms involving $\vec{u}$ are the usual hyperbolic terms and can be treated with the standard methods.
The last term on the right hand side can be implemented such that it
is an exact exchange term between the plasma and cosmic ray internal
energy equations; this term will probably need to be upwinded.  
The streaming term  (third term on the left hand side) is the
main subject of this paper, and will be examined using the following
conservative one-dimensional model equation describing streaming down
the gradient:
\be
\label{eq:basic}
\frac{\partial f}{\partial t} + \frac{\partial}{\partial x}(vf) = 0,
\ee
where $f(x,t)$ is a function of space and time, $v=-{\rm sgn}(\partial f/\partial x)$, and ${\rm sgn}(s)=1$ if
$s>0$, 0 if $s=0$, and $-1$ if $s<0$.  
The streaming equation is very similar in appearance to the simple 
advection equation (but very different in nature as we shall see), but
differs from a variable-coefficient advection equation as the
advection speed ($v$) depends, not on $x$, but  
on the gradient of $f$. From here on we focus on the simpler Eq. (\ref{eq:basic}) and effective numerical methods to solve it. Once numerical 
methods for solving Eq. (\ref{eq:basic}) are available, the streaming
part of Eq. (\ref{eq:pcr}) can be implemented in an analogous fashion.  

Integrating Eq. (\ref{eq:basic}) in space (from -$\Delta x/2$ to $\Delta x/2$) and time (from $-\Delta t/2$ to $\Delta t/2$) one obtains,
\be
\label{eq:control_vol}
\Delta x \left [  f_{0,0^+} - f_{0,0^-} \right ] + \Delta t \left [ \left . -f {\rm sgn}(\partial f/\partial x) \right |_{0^+,0} + \left .  f {\rm sgn}(\partial f/\partial x) \right |_{0^-,0}  \right ] = 0.
\ee
Evaluating Eq. (\ref{eq:control_vol}) at a maximum of $f$ (similar argument applies at a minimum), one obtains
$$
\Delta x \left [  f_{0,0^+} - f_{0,0^-} \right ] + \Delta t [ f_{0^+,0}+f_{0^-,0}] = 0,
$$
because the ${\rm sgn}$ function changes sign across the extremum. Assuming that $f$ is continuous in $x$, this reduces to 
$$ 
\Delta x  \left [  f_{0,0^+} - f_{0,0^-} \right ] + 2 \Delta t f_{0,0} =0. 
$$
Assuming that $f$ is also continuous in time, this is equivalent to $\Delta x \partial f/\partial t + 2 f =0$. In the limit $\Delta x \rightarrow 0$, this implies
$f=0$, an apparent contradiction. This contradiction also applies to the cosmic ray pressure equation (Eq. \ref{eq:pcr}; the last term on the right hand 
side of Eq. \ref{eq:pcr} reduces to zero on integrating in space at an extremum). A resolution of this apparent contradiction is that, rather than $f =0$,
$\partial f/\partial t \rightarrow -2f/\Delta x$; i.e., $f$ changes infinitely fast at an extremum. A similar behavior is observed for the diffusion equation
when applied at a point where $f$ is continuous but the gradient of $f$ changes discontinuously. 
The analogy with the diffusion equation is deeper; in section \ref{sec:reg} we 
show that the regularized equation is diffusive (and hence parabolic) in nature at extrema. 
For the diffusion equation the discontinuity in $f^\prime$ is removed instantly and the flux becomes continuous.
Similarly, when we 
regularize the ${\rm sgn}(f^\prime)$ 
function by $\tanh(f^\prime/\epsilon)$ ($f^\prime\equiv \partial f/\partial x$; see section \ref{sec:reg} for the details of regularization that we use), the 
flux $vf \equiv -f\tanh(f^\prime/\epsilon)$ instantly becomes continuous (see Fig. \ref{fig} for the plots of $vf$ at different times) at the extrema and 
$\partial f/\partial t$ is well behaved, just as it is for the diffusion equation. 
It is reassuring that a unique solution is obtained for the regularized equation (Eq. \ref{eq:reg1}) in the limit $\Delta x \rightarrow 0$ and 
$\epsilon \rightarrow 0$, for a sufficiently small timestep $\Delta t$. In Section \ref{sec:dif_reg} we show that a unique solution is obtained using 
other regularizations.

\comments{
Eq. (\ref{eq:basic}) can be expanded as
\be
\label{eq:expanded}
\frac{\partial f}{\partial t} + v \frac{\partial f}{\partial x} + f \frac{\partial v}{\partial x} = 0.
\ee
Eq. (\ref{eq:expanded}) reduces to the standard constant speed advection equation at all locations, except at the local extrema (and inflection points) of $f$,  as $\partial v/\partial x$= 0 at these locations. However, at extrema (and inflection points) $f^\prime=0$ ($f^\prime \equiv \partial f/\partial x$), but the numerical approximation of
$\partial v/\partial x=2/\Delta x$ at a maximum and $\partial v/\partial x=-2/\Delta x$ at a minimum ($\partial v/\partial x=0$ at an inflection 
point), where 
$\Delta x$ is the grid spacing. Thus, from Eq. (\ref{eq:expanded}) its easy to see that $f$ is
reduced at a maximum and increased at a minimum due to this large term ($f>0$ is assumed throughout the paper). 
The conservative form (Eq. [\ref{eq:basic}]) shows that 
at a maximum (minimum) fluxes in both directions move outward (inward), reducing (increasing) the local maximum (minimum). Thus, fluxes at extrema seem diffusive in nature; this we shall see explicitly with the regularized form in section (\ref{sec:reg}). Since Eq. 
(\ref{eq:basic}) is not hyperbolic at extrema, it is not hyperbolic in general. 
}

The methods for treating the standard hydrodynamic and magnetohydrodynamic (MHD) equations are quite mature by now, 
but entirely unexpected numerical problems arise   
when new physics (as in Eq. \ref{eq:pcr}) is added. Thus, the numerical implementation of these new terms must be tested rigorously. For example, simple centered 
finite differencing of anisotropic thermal conduction can give rise to negative temperatures for large temperature gradients, an unphysical result 
(\cite{Sharma2007}). Similarly, here we show that a naive implementation of Eq. (\ref{eq:basic}) results in unphysical oscillations.

The paper is organized as follows. Section (\ref{sec:non-mon}) shows that a standard explicit implementation of Eq. (\ref{eq:basic})
results in spurious oscillations of large amplitude. It is shown that a much smaller timestep with $\Delta t \propto \Delta x^3$ can remove oscillations
and results in a physically consistent evolution. In section (\ref{sec:reg}) we regularize Eq. (\ref{eq:basic}) and develop implicit methods based 
on it. Section (\ref{sec:test}) presents test problems (with both continuous and discontinuous initial profiles) which show that physically consistent solutions are obtained using implicit methods with a sufficiently small timestep $\Delta t \propto \Delta x$. In Section (\ref{sec:dif_reg}) we present two different
regularizations and show that the same solution is obtained in the converged limit. In section (\ref{sec:conc}) we discuss the nature of the solutions, and conclude with the implications of our methods for the fluid modeling of cosmic rays and for thermal conduction in the collisionless regime.

\section{Non-monotonicity with explicit schemes using the standard CFL time-step}
\label{sec:non-mon}
\begin{figure}
\begin{center}
\psfig{height=4.0in, figure=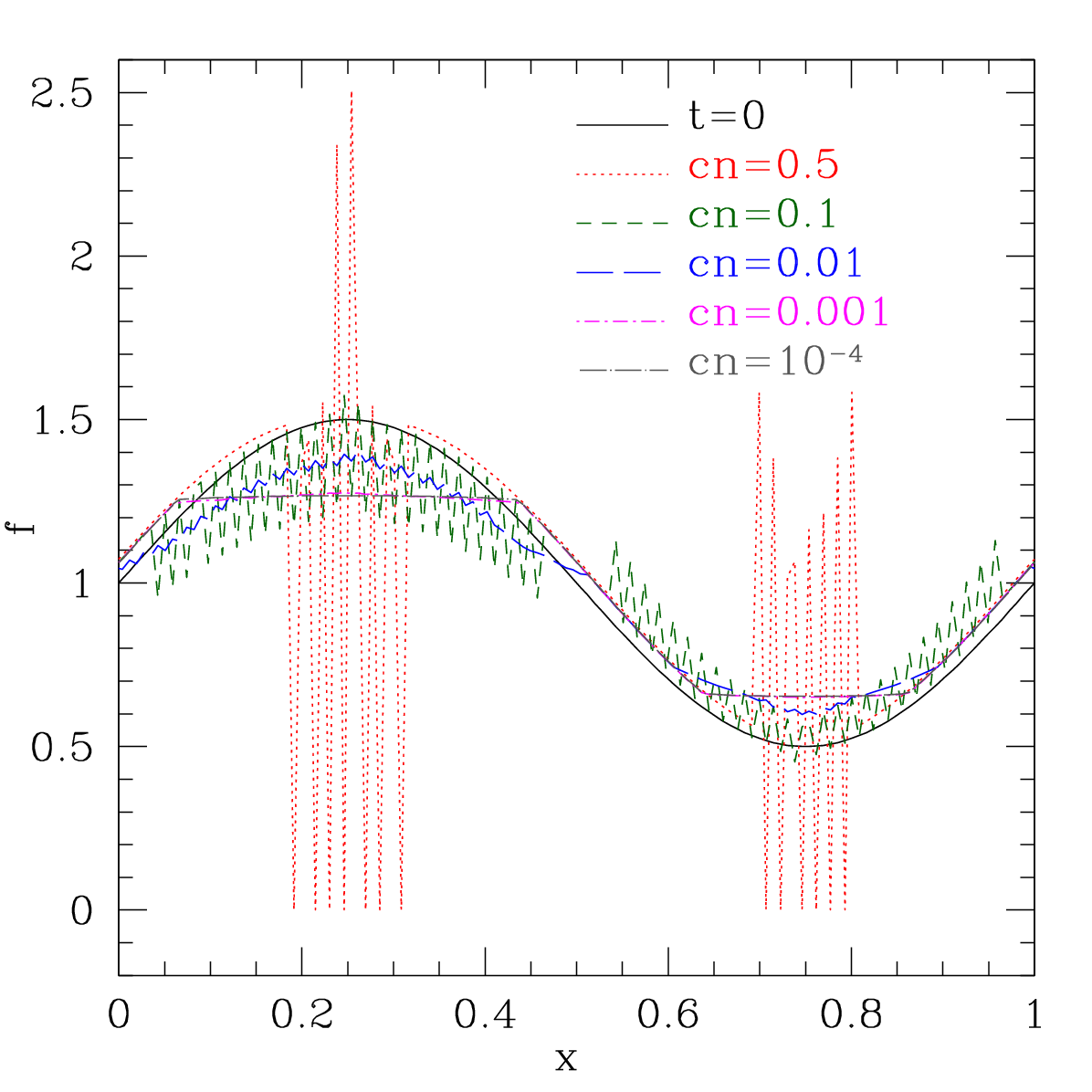}
\end{center}
\caption{Profiles at $t=0$ (solid black line) and $t=0.02$ for the sine wave initial condition (Eq. \ref{eq:sin}) with different CFL numbers. 
Number of grid points is 128. Periodic boundary conditions are used. The profiles with cn=0.001 and cn=10$^{-4}$ almost overlap.}
\label{fig:fig1}
\end{figure}
The standard methods for solving hyperbolic equations fail miserably
for solving Eq. (\ref{eq:basic}), as demonstrated by a periodic test 
problem with a smooth initial profile given by 
\be
\label{eq:sin}
f(x,0) = 1 + 0.5\sin(2\pi x).
\ee
We show that a standard explicit update scheme with a reasonable
Courant-Friedrichs-Lewy (CFL) number ($\Delta t= {\rm cn} \Delta
x/|v|$, where cn is the CFL number,  
$|v|=1$ for Eq. \ref{eq:basic}) gives rise to spurious oscillations
even for a smooth initial condition like Eq. (\ref{eq:sin}). Figure
(\ref{fig:fig1}) shows the initial profile and profiles at $t=0.02$
for runs with different CFL numbers; the number of grid points is
128. Eq. (\ref{eq:basic}) is evolved using an upwind method with a van
Leer slope limiter (e.g., see \cite{Leveque2002}); such oscillations
occur for all explicit methods. As mentioned in the introduction, the streaming 
equation (Eq. \ref{eq:basic}) is mathematically well-behaved only when the discontinuous 
${\rm sgn}(f^\prime)$ function is replaced by its smoothened form $\tanh(f^\prime/\epsilon)$; 
thus we use this smoothened form  with $\epsilon=0.1$ to calculate the fluxes. If we use the ${\rm sgn}$
function instead of the smoothened form, grid scale oscillations in $f$ arise even for an arbitrarily small $\Delta t$ 
(oscillation amplitude becomes smaller with a smaller $\Delta t$ though).
The profile with cn=0.5 shows large amplitude oscillations originating
from the initial extrema. Although the amplitude of oscillations with
cn=0.1 is smaller, the oscillations are more spread out; this is
because the number of timesteps to reach $t=0.02$ is larger with a smaller
CFL number (hence spreading out the oscillations).  As the CFL
number is reduced further the profiles converge to a reasonable-looking $f$. Profiles with cn=0.001 show small amplitude oscillations,
but the profile with cn=$10^{-4}$ does not.

\begin{figure}
\centering
\includegraphics[width=3in,height=3in]{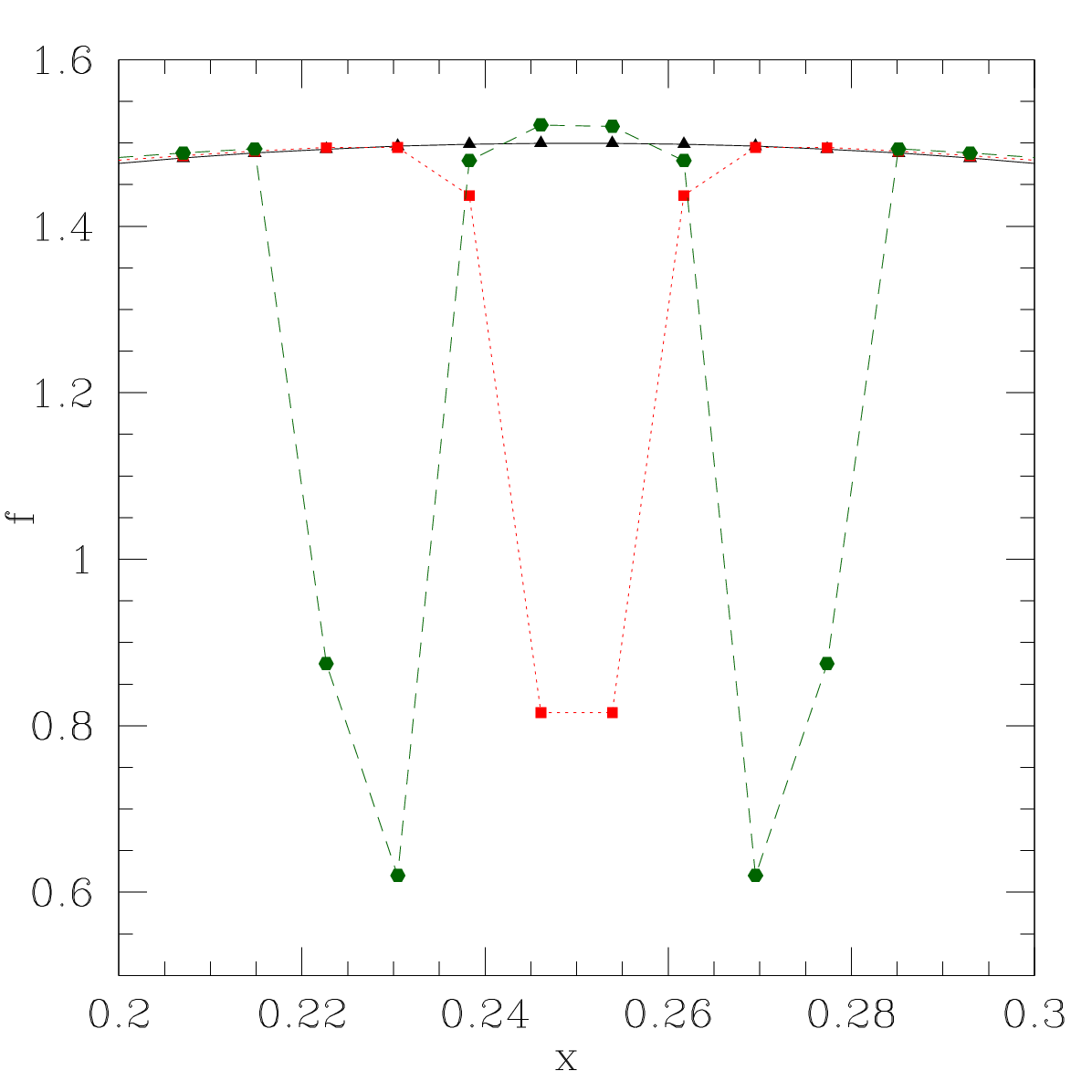}
\caption{Zoom in on the profiles at first two timesteps for cn=0.5 in Figure (\ref{fig:fig1}): the initial profile is marked by a solid line and triangles;  the 
profile at the first timestep by dotted line and squares; and the profile at the second timestep by  dashed line and circles. Grid-scale oscillations 
originate at the local extrema and propagate throughout the box with time.
\label{fig:fig2}}
\end{figure}

These oscillations originate from the local extrema. With a large timestep (even a CFL-limited timestep is large for the extrema), the fluxes (on 
both sides)
leaving the initial maximum  are big enough to reduce the initial maximum way below its initial value (see Fig. \ref{fig:fig2}), thus giving rise to 
an unphysical dip at the location of the initial maximum. Since the dip is a local minimum now, there are large fluxes entering those grid points in the next 
timestep, again making them a local maximum and simultaneously making the adjoining grid cells a minimum. Now two new local minima are created
in cells adjoining the initial maximum, in addition to the initial
maximum. In this way oscillations propagate 
away from the initial extremum as the solution is advanced, resulting
in oscillations which spread over the whole domain with time. 

\begin{figure}
\centering
\psfrag{A}{$f_0$}
\psfrag{B}{$f_{-1}$}
\psfrag{C}{$f_1$}
\psfrag{D}{$f_{-2}$}
\psfrag{E}{$f_2$}
\includegraphics[width=3in,height=2.2in]{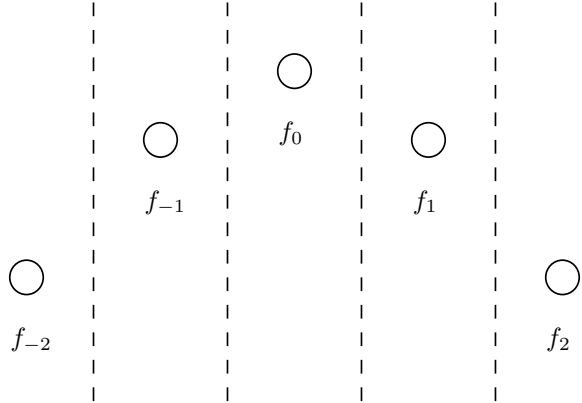}
\caption{A schematic representation of a maximum at $f_0$. Unphysically large fluxes leaving $f_0$ with the standard CFL-limited timestep result in oscillations seen in
Figs. (\ref{fig:fig1}) and (\ref{fig:fig2}).
\label{fig:fig3}}
\end{figure}

One can obtain a limit on the timestep such that new local extrema are not created. Consider a maximum (at $f_0$) shown schematically in Figure 
(\ref{fig:fig3}); we consider the extrema because the oscillations originate there. 
Considering a piecewise constant reconstruction, let us apply an upwind update of Eq. (\ref{eq:basic}) for a single timestep, such that, 
$f_0^{n+1}=f_0^n - 2 \Delta t/\Delta x f_0^n,~f_1^{n+1}=f_1^n + \Delta t/\Delta x (f_0^n -f_1^n)$. Now imposing monotonicity after a single timestep, i.e., 
$f_0^{n+1} \geq f_1^{n+1}$ gives,
 \be
 \label{eq:exp_mon}
 \Delta t \leq |f^{\prime\prime}| \Delta x^3/(4|v|f), 
 \ee
 since $f_0^n-f_1^n\approx |f^{\prime\prime}|\Delta x^2/2$ and $f_0^n \gg 
|f^{\prime\prime}| \Delta x^2$, where $f^{\prime\prime}$ ($\equiv \partial^2 f/\partial x^2$) is evaluated at the extremum. Notice that the timestep restriction will 
be worse if even higher order derivatives vanish, and also that  $|f^{\prime\prime}|$ decreases as the initial profile smoothens with time. 
A similar condition for monotonicity is obtained at a minimum.
Although we have considered an upwind method above, a similar timestep constraint is
obtained for any explicit method.
A result similar to Eq. (\ref{eq:exp_mon}) is obtained by analyzing the regularized form of Eq. (\ref{eq:basic}) in section (\ref{sec:reg}).
The value of $|f^{\prime \prime}|$ at extrema for the initial profile in Eq. (\ref{eq:sin}) is $2\pi^2$ and the limit on timestep given by Eq. (\ref{eq:exp_mon})
is $ \Delta t \leq (\pi^2\Delta x^2/3) \Delta x$, which on using $\Delta x=1/128$ gives $\Delta t \leq 2 \times 10^{-4} \Delta x$. This is consistent with 
oscillations seen in Figure (\ref{fig:fig1}) for cn$\geq0.001$ and not for cn=10$^{-4}$.

\section{Regularization}
\label{sec:reg}

Since the advection velocity $v$ can change discontinuously at extrema, the differential form in Eq. (\ref{eq:basic}) is well posed only in its regularized
form. A similar situation occurs for the Euler/Burger's equations in the absence of viscosity. Motivated by the use of explicit viscosity to make Euler/Burger's equations well-posed, we introduce a physically motivated regularization of Eq. (\ref{eq:basic}). A straightforward regularization is obtained by replacing 
sgn$(f^\prime)$ in Eq. (\ref{eq:basic}) by its smooth approximation $\tanh(f^\prime/\epsilon)$, where $\epsilon$ is a small parameter.

Thus the regularized form of Eq. (\ref{eq:basic}) becomes,
\ba
\label{eq:reg1}
\frac{\partial f}{\partial t} & - & \frac{\partial}{\partial x} \left [ f \tanh(f^\prime/\epsilon) \right ] = 0, \hspace{1.5in} {\rm or} \\
\label{eq:reg}
\frac{\partial f}{\partial t} &-& \tanh(f^\prime/\epsilon) \frac{\partial f}{\partial x} =  \frac{f}{\epsilon} {\rm sech}^2(f^\prime/\epsilon) \frac{\partial^2 f}{\partial x^2},
\ea
where we have used $d\tanh(s) = {\rm sech}^2(s)ds$. While the second term on the left hand side of Eq. (\ref{eq:reg}) is an advective term,
the term on the right hand side of Eq. (\ref{eq:reg}) is a diffusive term, which at an extremum (where $f^\prime=0$) is $ff^{\prime\prime}/\epsilon$, a 
diffusive term with the diffusion coefficient $f/\epsilon$. There is a physical justification for such a regularization. As mentioned before, away from the
cosmic ray pressure extrema, cosmic ray fluid streams down its pressure gradient along magnetic field lines; in addition to this streaming there is
also diffusion (due to pitch angle scattering by the self-excited Alfv\'en waves, which is typically small, but should be added on the right hand side of 
Eq. [\ref{eq:pcr}]) along magnetic field lines.  At cosmic ray pressure extrema, the cosmic ray particles have an equal probability to move in either direction
(along magnetic field lines) at the speed of light. However, they are scattered after traveling a distance of the order of the Larmor radius, thus resulting in a diffusive behavior at extrema. Just as viscosity can be ignored at all locations except close to shocks for Euler/Burger's equations, the diffusive behavior of Eq. (\ref{eq:reg}) is only effective at extrema.

The time step limit for the stability of the diffusive term in Eq. (\ref{eq:reg}) treated explicitly is 
\be
\label{eq:stab}
\Delta t \leq \Delta x^2 \epsilon/2f; 
\ee
on taking $\epsilon \sim f \Delta x/L^2$ ($L$ is the box-size), this scales in the same way as Eq. (\ref{eq:exp_mon}). Although this restriction
on timestep follows from the linear stability analysis of the discretized diffusion equation, explicit methods with larger timesteps do not blow up because
the large diffusion is only concentrated in an infinitesimally small region where $f^{\prime\prime}\approx0$. Away from extrema, 
the term on the left hand side is like advection. Eq. (\ref{eq:reg}) clearly shows that Eq. (\ref{eq:basic}) is not hyperbolic, and because of that the methods
for hyperbolic equations do not work for this. Since the regularized form in Eq. (\ref{eq:reg1}) is more transparent and has a less restrictive timestep requirement, 
we will work with it for testing different explicit and implicit methods. We only focus on conservative schemes because the underlying equation (Eq. \ref{eq:basic}) is conservative.

\subsection{Explicit methods}
Explicit methods are not as attractive as implicit methods for this problem because of the restrictive timestep (Eq. [\ref{eq:stab}]) for stability, but we discuss them 
for completeness. With the standard CFL timestep, both the Lax-Wendroff and upwind methods give very similar large amplitude oscillations,
even with a smooth initial profile (see Fig. [\ref{fig:fig1}]). With a smaller timestep scaling like $\Delta x^2$ (for a fixed $\epsilon$) both methods give
physically consistent results.

\subsubsection{Lax-Wendroff method}
The second order accurate Lax-Wendroff discretization (\cite{Lax1960}) for Eq. (\ref{eq:reg1}) is obtained by observing that
$$
f(x,t+\Delta t) = f(x,t) + \Delta t \frac{\partial f}{\partial t} + \frac{\Delta t^2}{2} \frac{\partial^2 f}{\partial t^2},
$$
where partial derivatives ($\dot{f} \equiv \partial f/\partial t$) are evaluated at $(x,t)$. Using, $\dot{f}= [f \tanh(f^\prime/\epsilon)]^\prime $ (Eq. [\ref{eq:reg1}]), 
$\ddot{f}=[ \partial(f \tanh(f^\prime/\epsilon))/\partial t ]^\prime$, and expanding the time derivatives in terms of space derivatives,
one gets,
\be
\frac{f^{n+1}_i-f^n_i}{\Delta t} = \frac{\partial}{\partial x} \left \{  f \tanh(f^\prime/\epsilon) + \frac{\Delta t}{2} \left [ \tanh(f^\prime/\epsilon)[f \tanh(f^\prime/\epsilon)]^\prime + \frac{f}{\epsilon} {\rm sech}^2(f^\prime/\epsilon) [f \tanh(f^\prime/\epsilon)]^{\prime \prime}   \right ]  \right \}, 
\ee
where quantities on the right hand side are simply averaged to be centered at $(i, n)$. The above discretization is clearly conservative. 

\subsubsection{Upwind method}
The explicit upwind discretization of Eq. (\ref{eq:reg1}) is given by
\be
f^{n+1}_i =  f_i^n - \frac{\Delta t}{\Delta x} \left (  F_{i+1/2}^{n+1/2} - F_{i-1/2}^{n+1/2} \right ),
\ee
where the flux ($F_{i+1/2}^{n+1/2}$) is calculated in an upwind fashion using a standard first order slope limiter  (we are using the van Leer limiter; \cite{vanLeer1977,Leveque2002}). 
The flux equals
\be
\nonumber
F_{i+1/2}^{n+1/2} = 
\cases {
v  \left ( f_i^n + \sigma_i \frac{[\Delta x - v\Delta t]}{2} \right ),
&{\rm for}$~v>0$, \cr
\nonumber
 v  \left ( f_{i+1}^n - \sigma_{i+1} \frac{[\Delta x + v\Delta
    t]}{2}  \right ), & {\rm for} $~v\leq 0$,\cr
 }
\ee
where $v=-\tanh(f^\prime/\epsilon)_{i+1/2}^n$, $\sigma_i = L[f^\prime_{i+1/2},f^\prime_{i-1/2}]$ ($L$ is a slope limiter like van Leer's).

\subsection{Implicit methods}
Implicit methods are attractive because longer timesteps, scaling as $\Delta x$ and not  as $\Delta x^2$, can be used to give stable and converging results as the resolution is increased.
\subsubsection{Linearized implicit method}
Assuming smoothness of $f^\prime$, we can write the backward-Euler expansion of  $\tanh(f^\prime/\epsilon)$ evaluated at timestep $(n+1)$ in 
Eq. (\ref{eq:reg1}) as
\be
\label{eq:lin_exp}
\tanh(f^\prime/\epsilon)^{n+1} =  \tanh(f^\prime/\epsilon)^n + \frac{1}{\epsilon}{\rm sech}^2(f^\prime/\epsilon)^n \left ( f^{\prime,n+1} - f^{\prime,n} \right).
\ee
In writing above we have linearized the implicit approximation  in $f^{\prime,n+1}$. The linearized (semi-)implicit flux for the discretized equation (centered in 
space), using above, can be written as
\ba
\label{eq:imp_lin}
F^{n+1}_{i+1/2} = - & \Bigl( & f_{i+1}^{n+1} \left [ \frac{1}{2}
    \tanh(f^\prime/\epsilon)^n_{i+1/2} + \frac{1}{\epsilon \Delta x}
    {\rm sech}^2(f^\prime/\epsilon)_{i+1/2}^n f_i^n \right ] \\
 & + & f_i^{n+1} \left [ \frac{1}{2} \tanh(f^\prime/\epsilon)^n_{i+1/2}  -
   \frac{1}{\epsilon \Delta x} {\rm
     sech}^2(f^\prime/\epsilon)_{i+1/2}^n f_{i+1}^n \right] \Bigr), \nonumber
\ea
which is linear in $f^{n+1}$. Thus the implicit form of Eq. (\ref{eq:reg1}), using above,  can be evolved using the standard tridiagonal solver (since we are imposing
periodic boundary conditions the tridiagonal solver is combined with the Sherman-Morrison formula; e.g., \cite{Press1992}).

\subsubsection{Nonlinear implicit method}
Instead of linearizing the implicit equation, as we did in the previous section, we can use a fully-implicit nonlinear method. The nonlinear implicit
equation can be solved using a Krylov subspace method like the generalized minimal residual
method (GMRES; \cite{Saad1986}).
 Eq. (\ref{eq:reg1}) can be discretized as a nonlinear matrix equation $M(f^{n+1})f^{n+1}=f^n$, where 
the matrix-vector product to be used in GMRES is given by 
\be
MV_i = V_i - \frac{\Delta t}{\Delta x} \left \{  V_{i+1/2} \tanh(V^\prime/\epsilon)_{i+1/2} -  V_{i-1/2} \tanh(V^\prime/\epsilon)_{i-1/2} \right \}.
\ee
The matrix equation is solved iteratively until the relative error is small enough (we choose $10^{-12}$ as the relative error tolerance).
Since matrix $M$ is diagonally dominant, for a sufficiently small $\Delta t$, this implicit scheme converges rapidly.

\subsubsection{Semi-implicit method}

The implicit methods we consider are formally only first order accurate in time, but can be made second order accurate by combining implicit and 
explicit methods, similar to the Crank-Nicolson scheme for the diffusion equation. The regularized equation (Eq. [\ref{eq:reg1}]) can be discretized 
as
\ba
\nonumber
\frac{f^{n+1}_i-f^n_i}{\Delta t} &=& \frac{g}{\Delta x} \left [ f_{i+1/2}\tanh(f^\prime/\epsilon)_{i+1/2} - f_{i-1/2} \tanh(f^\prime/\epsilon)_{i-1/2} \right ]^{n}\\
\label{eq:vnsa}
&+& \frac{(1-g)}{\Delta x}  \left [ f_{i+1/2}\tanh(f^\prime/\epsilon)_{i+1/2} - f_{i-1/2} \tanh(f^\prime/\epsilon)_{i-1/2} \right ]^{n+1},
\ea
where the update is split into an explicit and implicit part. We update the explicit piece using the Lax-Wendroff method and the implicit piece using
the nonlinear implicit scheme. Results for different test problems are similar as the implicit schemes. Somewhat surprisingly, the convergence with 
increasing number of grid points for the semi-implicit ($g=1/2$ in Eq. [\ref{eq:vnsa}]) scheme is not very different from the fully implicit methods 
(see Fig. [\ref{fig:fig6}]). Notice that $f^\prime$ is not continuous (discontinuity occurs where the flattened maxima/minima is connected to the 
advecting part of the solution; e.g., Figs. [\ref{fig:fig4}a, \ref{fig:fig5}a]) for the streaming sine wave profile so the order of convergence is not easy to
establish analytically. Also, the timestep is so small that the temporal error is dominated by the spatial truncation error.

\section{Test problems}
\label{sec:test}
In this section we compare the different methods from section (\ref{sec:reg}), as they are applied to solve the regularized equation (Eq. [\ref{eq:reg1}]). We consider 
two periodic, one-dimensional test problems: an initial sine wave given by Eq. (\ref{eq:sin}) and an initial square wave. We use $\epsilon = 0.1$ 
unless mentioned otherwise.
\subsection{Smooth initial profile}
\begin{figure}
\begin{center}
\subfigure[$\Delta t = \epsilon \Delta x^2/3$]{\psfig{width=2.5in,figure=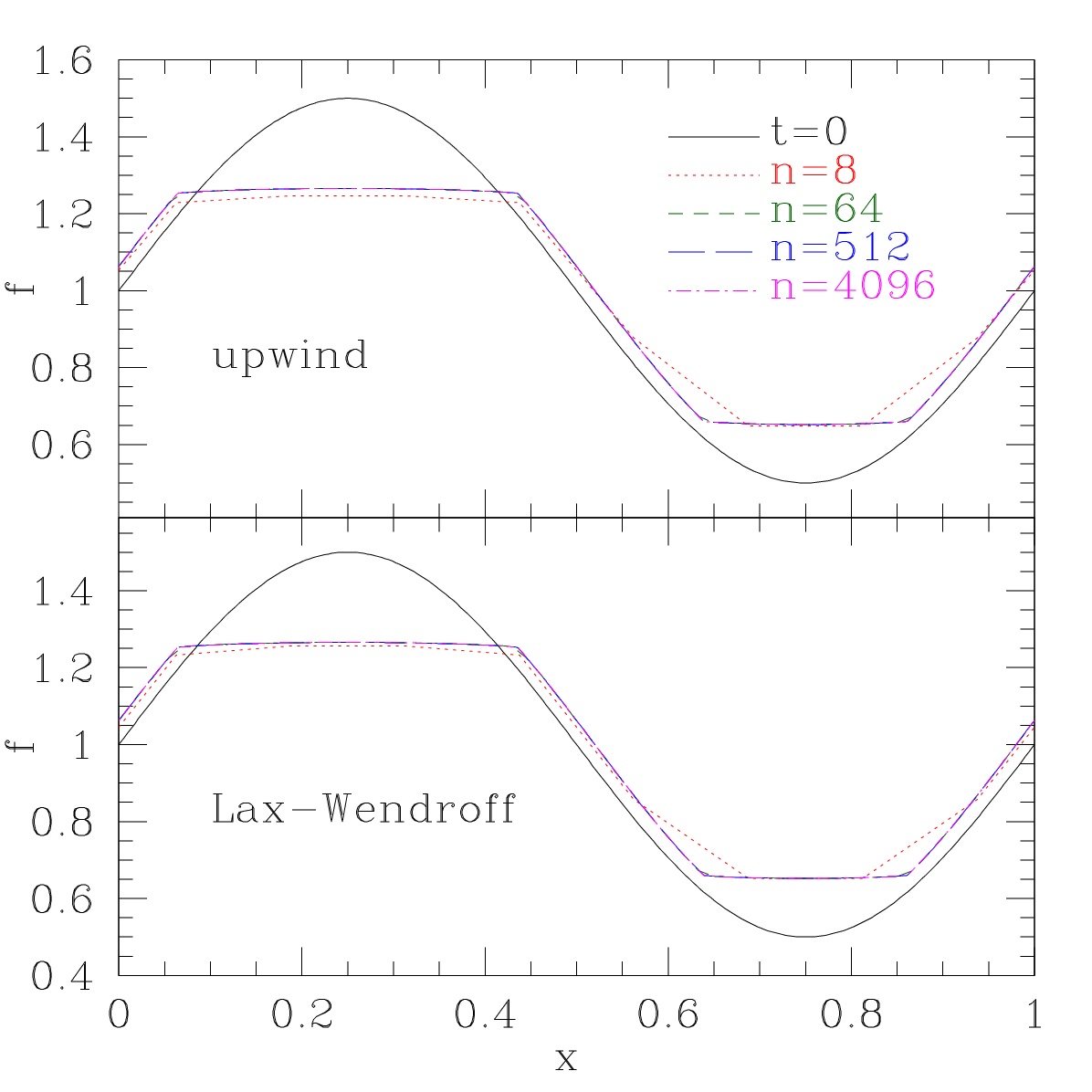}}
\subfigure[$\Delta t = 0.002 \Delta x$]{\psfig{width=2.5in,figure=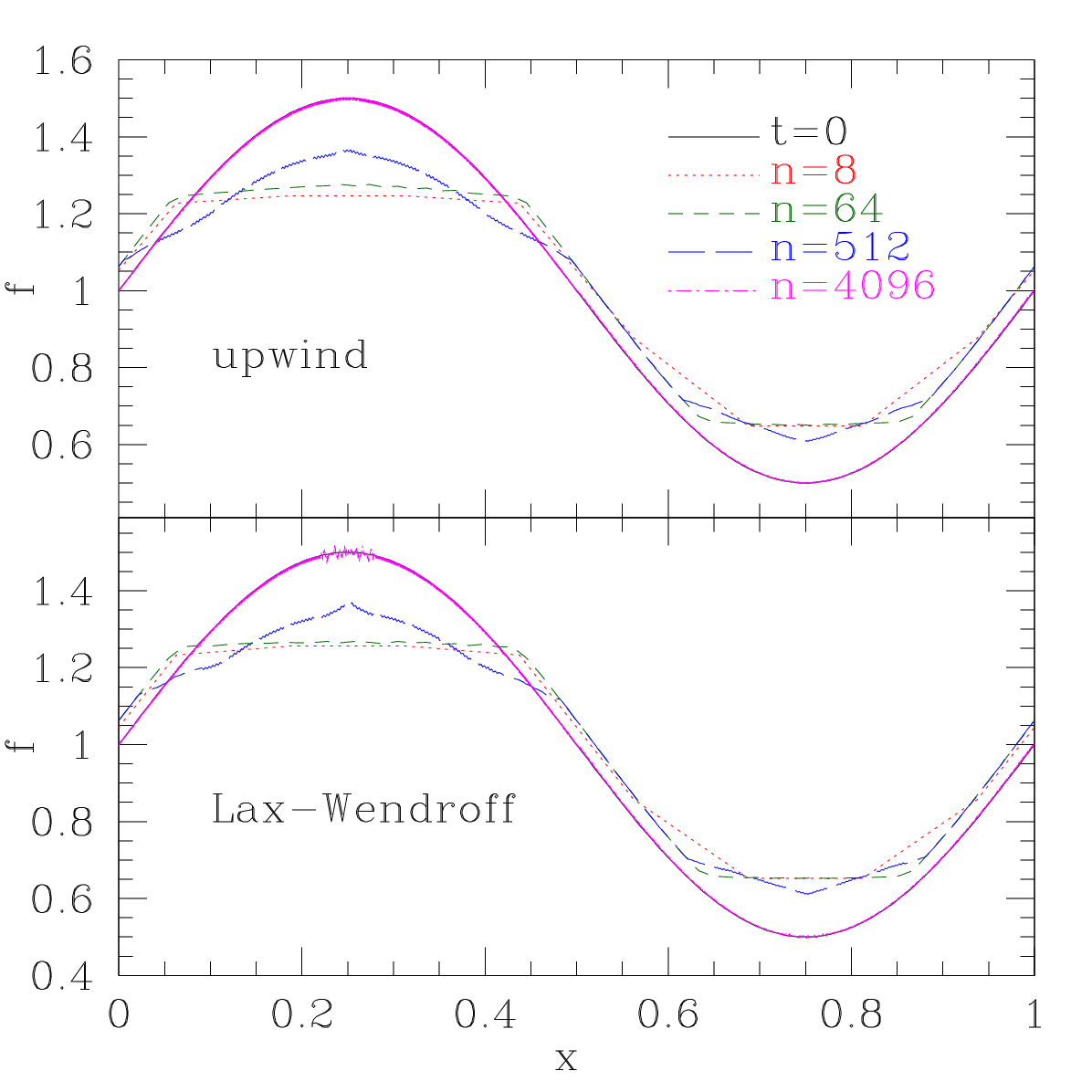}}
\end{center}
\caption{The initial profile (Eq. [\ref{eq:sin}]) and profiles at
  $t=0.02$ using (a) $\Delta t = \epsilon \Delta x^2/3$  ($\epsilon=0.1$) and (b) $\Delta t = 0.002 \Delta x$  for 
explicit schemes. Results with upwind and Lax-Wendroff methods look similar. Unlike with $\Delta t \propto \Delta x^2$, with $\Delta t \propto \Delta x$ the 
results do not converge as the spatial resolution is increased.}
\label{fig:fig4}
\end{figure}
Figure (\ref{fig:fig4}) shows profiles at $t=0.02$, beginning with the initial profile in Eq. (\ref{eq:sin}), for explicit methods using $\Delta t = \epsilon \Delta x^2/3$ Fig (\ref{fig:fig4}a; 
consistent with the timestep limit in Eq. [\ref{eq:stab}]) and $\Delta t = 0.002 \Delta x$ (Fig [\ref{fig:fig4}b]). While profiles converge to the physically consistent solution
with increasing resolution when $\Delta t \propto \Delta x^2$, increasing resolution with $\Delta t \propto \Delta x$ results in unphysical results as the resolution is increased. 
The convergence of $\Delta t \propto \Delta x^2$  and the non-convergence of $\Delta t \propto \Delta x$ is seen clearly from the plot in Figure (\ref{fig:fig5a}a). 
While $L_1$ error decreases for $\Delta t \propto \Delta x^2$, it increases for number of grid points larger than 16 for $\Delta t = 0.002 \Delta x$. Non-convergence
starts  once the timestep is longer than the limit in Eq. (\ref{eq:stab}), which happens for $n>16$ for $\Delta t=0.002 \Delta x$.
Although Eq. (\ref{eq:stab}) is the stability time limit on the diffusion term in Eq. (\ref{eq:reg}), results do not blow up because diffusion $\propto 
1/\epsilon$  is only limited to a small length-scale $\propto \epsilon$.

\begin{figure}
\begin{center}
\subfigure[implicit methods, $\epsilon=0.1$]{\psfig{width=2.5in, figure=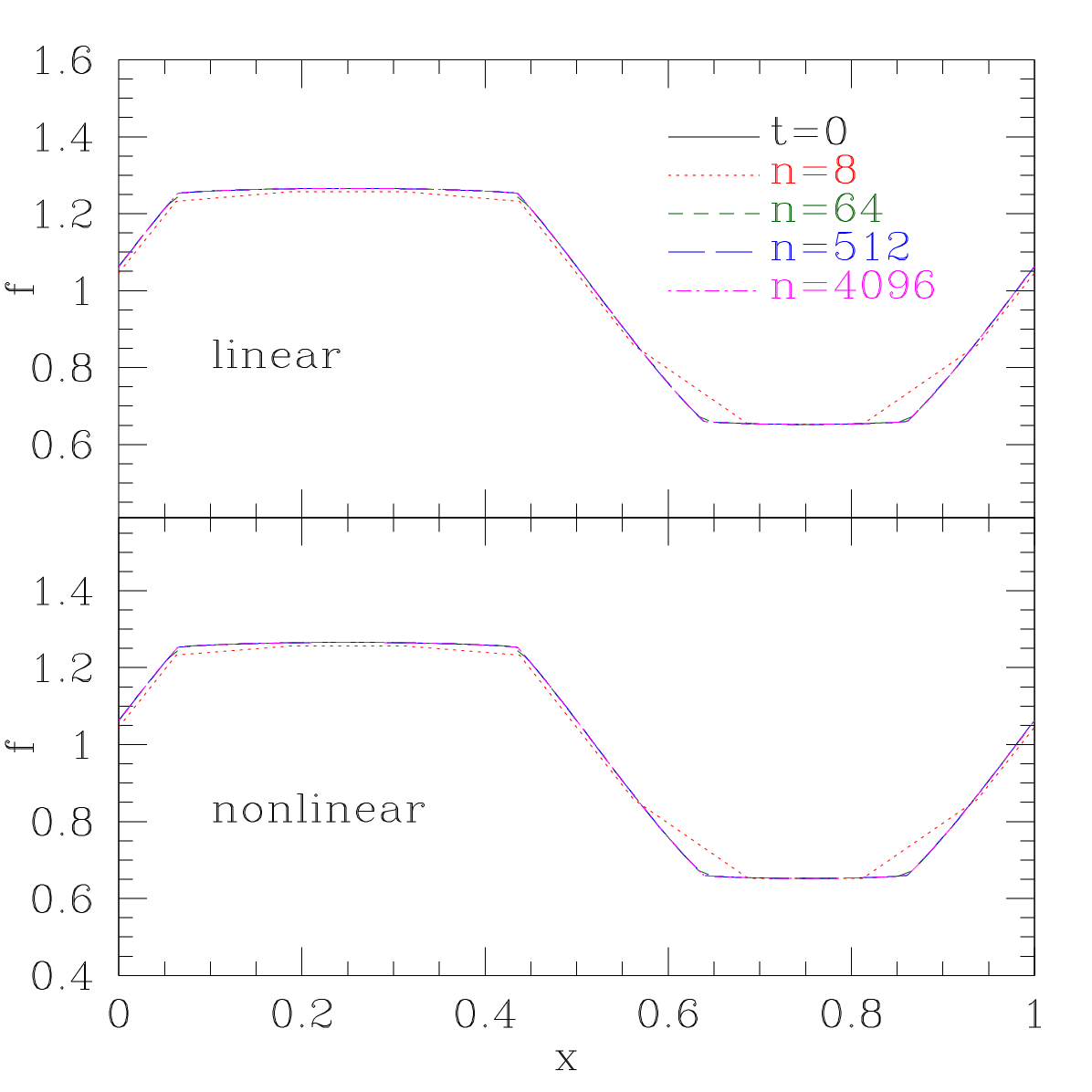}}
\subfigure[linearized implicit method, $\epsilon=0.01$]{\psfig{width=2.5in, figure=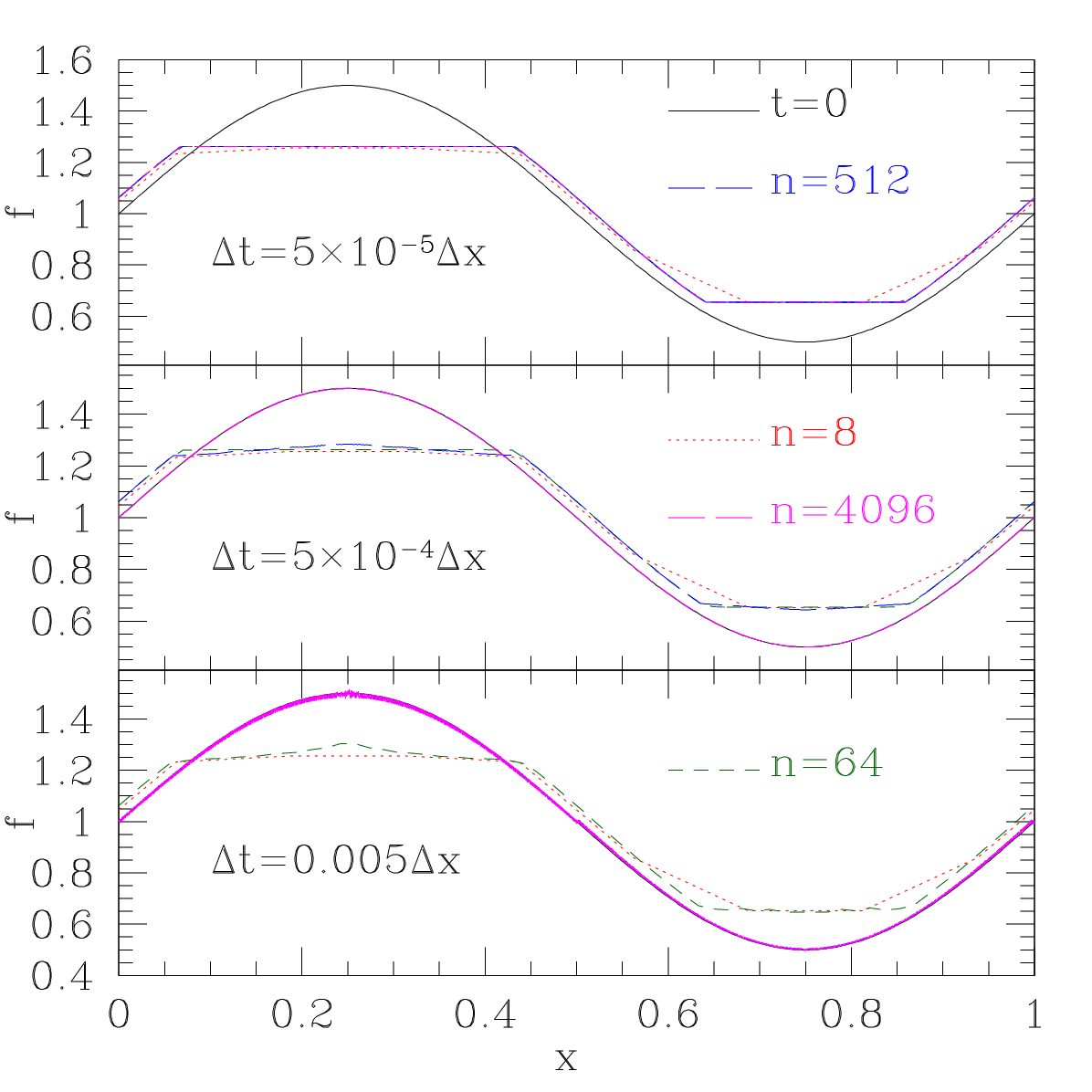}}
\end{center}
\caption{The initial profile (Eq. [\ref{eq:sin}]) and profiles at
  $t=0.02$ using  $\Delta t = 0.005 \Delta x$  for (a) implicit
  schemes with $\epsilon=0.1$ , and (b) the linearized implicit method
  for $\epsilon=0.01$ and with $\Delta t=0.005\Delta
  x,~5\times10^{-4}\Delta x$, $5\times10^{-5}\Delta x$.  Results with
  the linearized implicit method and the nonlinear implicit method
  look similar. Even for implicit methods, converged results are
  obtained only when $\Delta t$ is sufficiently small.} 
\label{fig:fig5}
\end{figure}

Quite surprisingly, the numerical experiments show that the implicit methods also require a small enough $\Delta t$ 
for convergence. This is probably because of the need to accurately resolve diffusion over
small length scales ($\propto \epsilon$). With $\epsilon=0.1$, only for $\Delta t \leq 0.005 \Delta x$ do the implicit methods give non-oscillatory convergent 
results (we have verified this for the number of grid points up to $n=262144=2^{18}$). Figure (\ref{fig:fig5}a) shows profiles at $t=0.02$ with implicit methods for the sine wave problem for $\epsilon=0.1$ and $\Delta t=0.005\Delta x$.  One gets converged results for $\Delta t \propto \Delta x$, unlike with the explicit methods (it is unclear if it 
holds as $\Delta x \rightarrow 0$ or if non-convergence appears for number of grid points more than the maximum we have tried). 

Figure (\ref{fig:fig5}b) shows profiles for $\epsilon=0.01$ but
with three different choices for  $\Delta t$, using the linearized implicit method (the nonlinear implicit method gives similar results). Using the same $\Delta t$ that gave converged results for $\epsilon=0.1$ ($\Delta t=0.005\Delta x$), one does not get 
converged results for $n>16$ (see Fig. [\ref{fig:fig5a}b]). Even using a ten times smaller timestep ($\Delta t = 5\times 10^{-4}\Delta x$; naively one would expect the stable timestep limit to scale with $\epsilon$ because at extrema, where oscillations originate, $\tanh[f^\prime/\epsilon] \approx f^\prime/\epsilon$) results in non-convergence for $n>64$ (see Fig. [\ref{fig:fig5a}b]). However, with $\Delta t = 5\times 10^{-5}\Delta x$ one can see convergence until $n=4096$ as seen in Figures (\ref{fig:fig5}b) \& (\ref{fig:fig5a}b); we have verified convergence until $n=32768=2^{15}$ for this case. 
Although the plot in Figure (\ref{fig:fig5}b) was done with the
linearized implicit method, a similar conclusion is reached for the
nonlinear implicit method; namely, the nonlinear iterative method
converges only if $\Delta t$ is similar to the value of $\Delta t$
which gave converged results with the linearized implicit method .  

\begin{figure}
\begin{center}
\subfigure[explicit methods with $\epsilon=0.1$]{\psfig{width=2.5in, figure=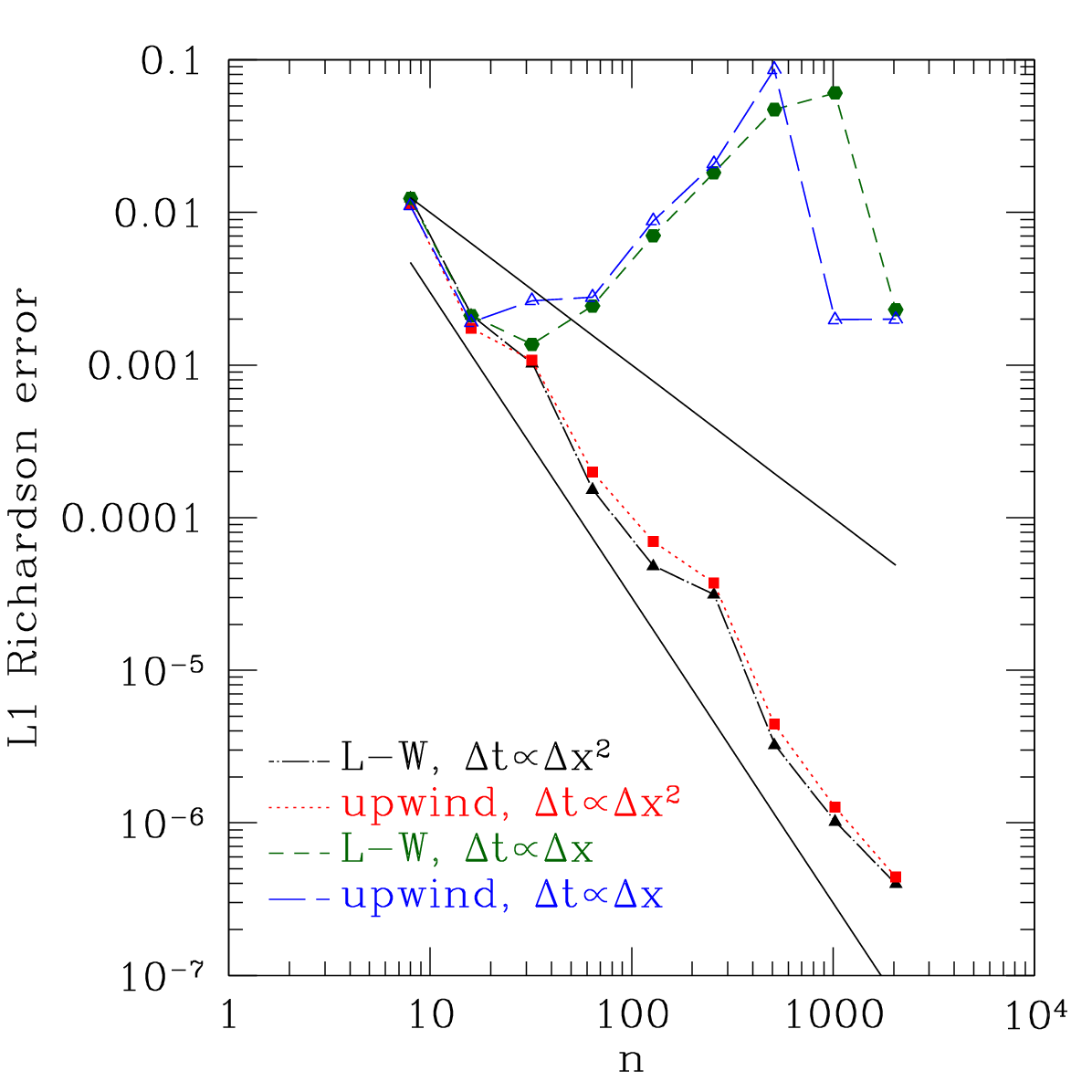}}
\subfigure[linearized implicit method with $\epsilon =0.01$]{\psfig{width=2.5in, figure=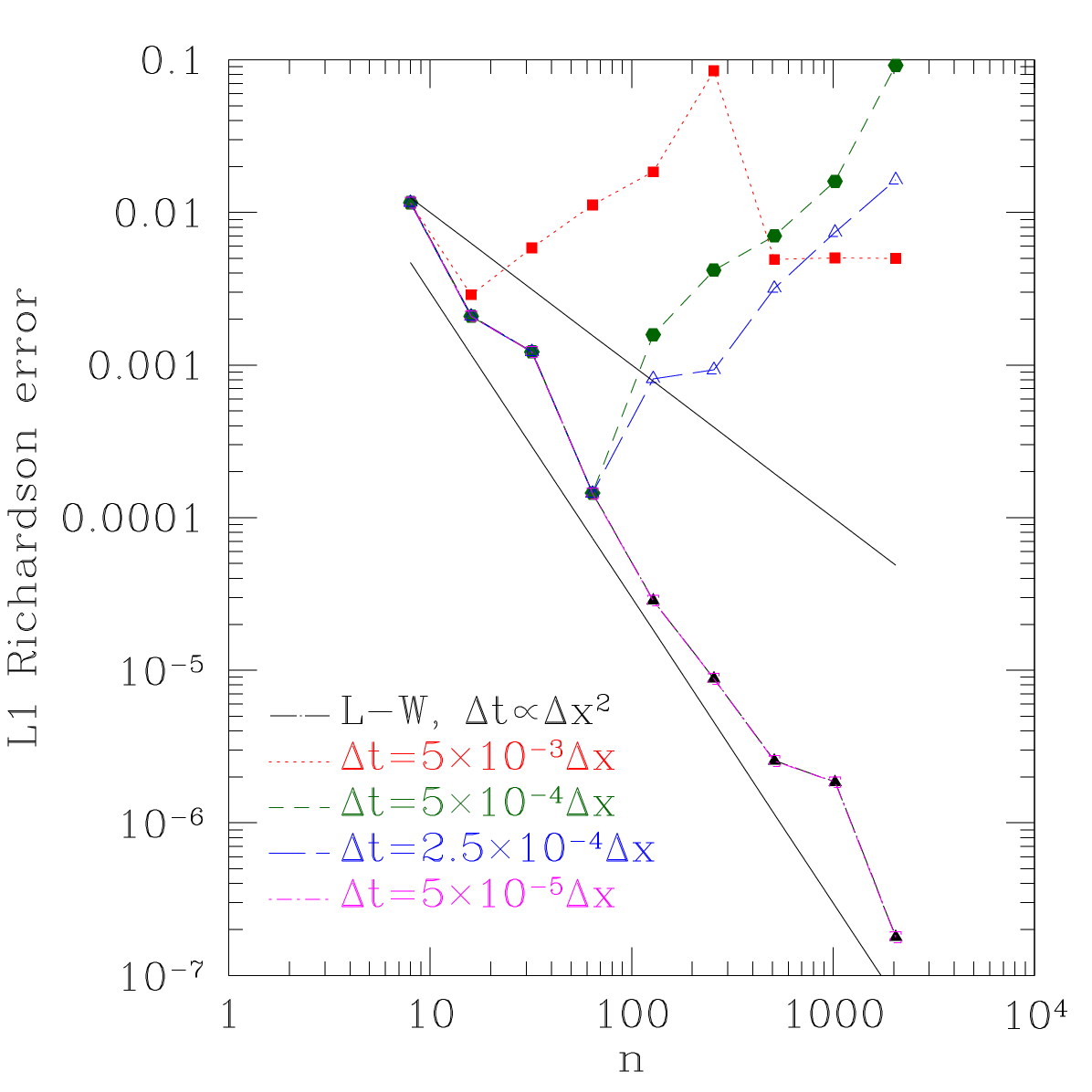}}
\end{center}
\caption{$L_1$ Richardson error (at $t=0.02$ for the initial profile
  in Eq. [\ref{eq:sin}]) as a function of number of grid points for
  (a) explicit methods with $\epsilon=0.1$ (corresponding to
  Fig. [\ref{fig:fig4}]) and  (b) the linearized implicit with $\epsilon = 0.01$ (corresponding to Fig. [\ref{fig:fig5}b]) method using 
different $\Delta t$. Timesteps are chosen to be
$\Delta t=\epsilon \Delta x^2/3$ and $\Delta t = 0.002 \Delta x$ for the explicit methods, and $\Delta t \propto \Delta x$ for the implicit schemes. As seen in Figs.
(\ref{fig:fig4}) and (\ref{fig:fig5}) converging results are obtained
only for a sufficiently small $\Delta t$. First- and second-order
convergence is indicated by solid lines. }
\label{fig:fig5a}
\end{figure}

\subsubsection{Numerical Convergence}

\begin{figure}
\centering
\includegraphics[width=3in,height=3in]{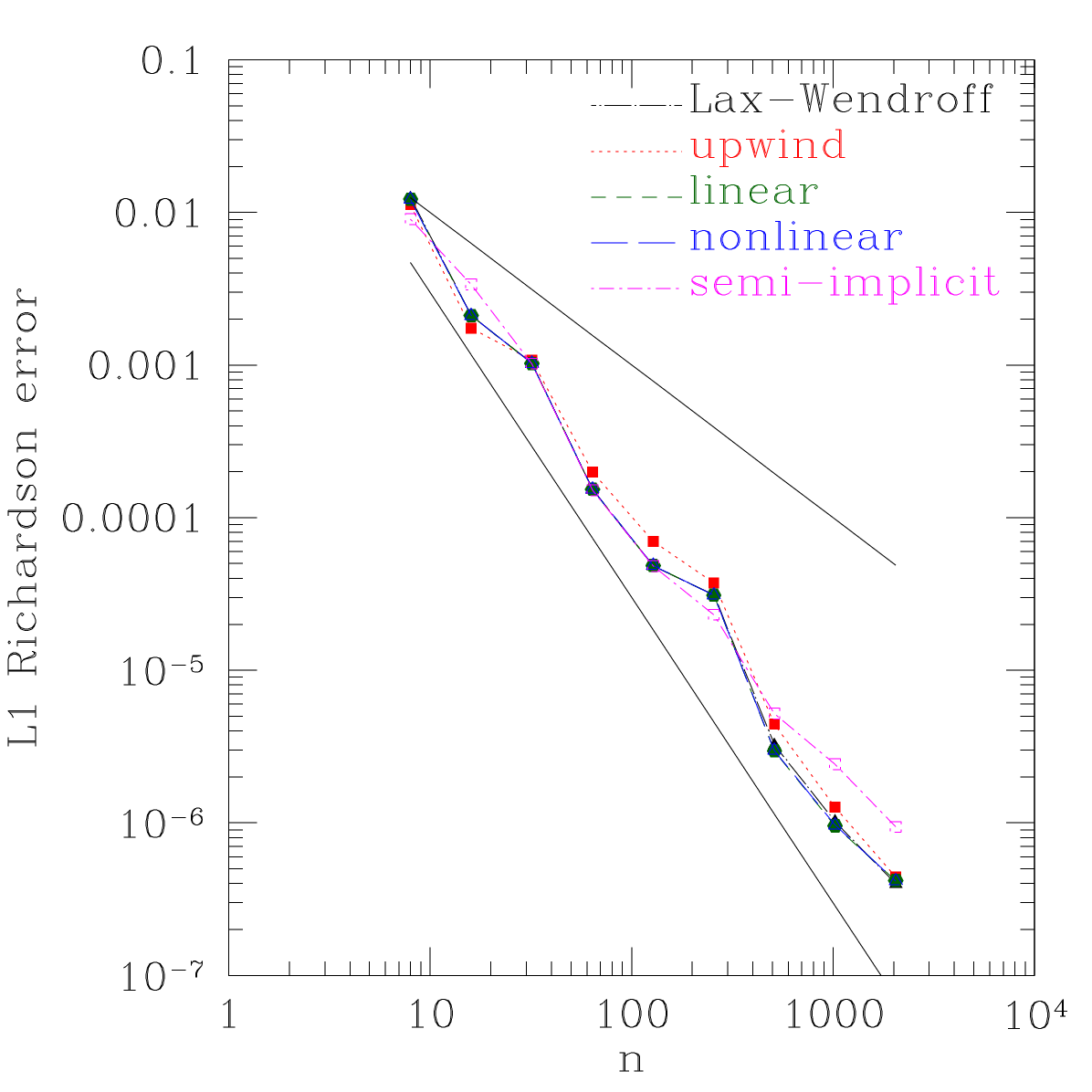}
\caption{$L_1$ Richardson error as a function of number of grid points at $t=0.02$ with different methods for the initial profile in Eq. (\ref{eq:sin}). 
The errors for Lax-Wendroff, linearized implicit, and nonlinear implicit methods lie almost on top of each other. The solid black lines show scalings as 
$n^{-1}$ and $n^{-2}$. 
\label{fig:fig6}}
\end{figure}

Figures (\ref{fig:fig5a}) \& (\ref{fig:fig6}) show the error (for the solution at $t=0.2$) for different methods with the initial sine wave test problem (Eq. [\ref{eq:sin}]);
Figure (\ref{fig:fig5a}a) shows errors with explicit methods (using $\epsilon=0.1$) for $\Delta t \propto \Delta x^2$ and $\Delta t \propto \Delta x$; 
Figure (\ref{fig:fig5a}b) shows errors with implicit methods (using $\epsilon=0.01$) for different $\Delta t$ scalings; Figure (\ref{fig:fig6})
shows errors using $\epsilon=0.1$, with a stable timestep, for all the different methods discussed in this paper. Since  the 
analytic solution is unknown, we show the Richardson errors. The $L_1$ Richardson error is given by 
$\Sigma_{i=1}^n |f_i - \overline{f}_i|/n$, 
where $f_i$ is the numerical solution for $f$, with $n$ grid points, at the $i^{\rm th}$ grid point, and $\overline{f}_i$ is the 
interpolation of the solution with $2n$ grid points at the same location.  Figure (\ref{fig:fig5a}) clearly shows that only for a sufficiently small $\Delta t$ do
the different (both explicit and implicit) methods converge.

Figure (\ref{fig:fig6}) shows that the convergence properties of different methods, using a stable timestep, are quite similar; in fact, the errors
lie almost on top of each other for upwind, linearized implicit, and fully nonlinear methods. All methods show a close to second-order convergence in the
$L_1$ norm. Somewhat surprisingly the $L_1$ error at the highest resolution is maximum for the semi-implicit method which, unlike other schemes, is formally
second order in time (however, $L_2$ errors for the semi-implicit method were smaller at the maximum resolution). As noted before, since $f^\prime$ is
discontinuous for the solution, a simple order of convergence does not apply.

\comments{
\begin{figure}
\begin{center}
\psfig{height=4.0in, figure=profiles.eps}
\end{center}

\caption{Converged profiles of $f$ at different times for the sine wave test problem (Eq. \ref{eq:sin}) using the linearized implicit method with $\epsilon=0.1$, $\Delta t=0.005 \Delta x$. Number of grid points is 2048.}
\label{fig1}
\end{figure}
}

\begin{figure}
\begin{center}
\subfigure[]{\psfig{width=2.5in, figure=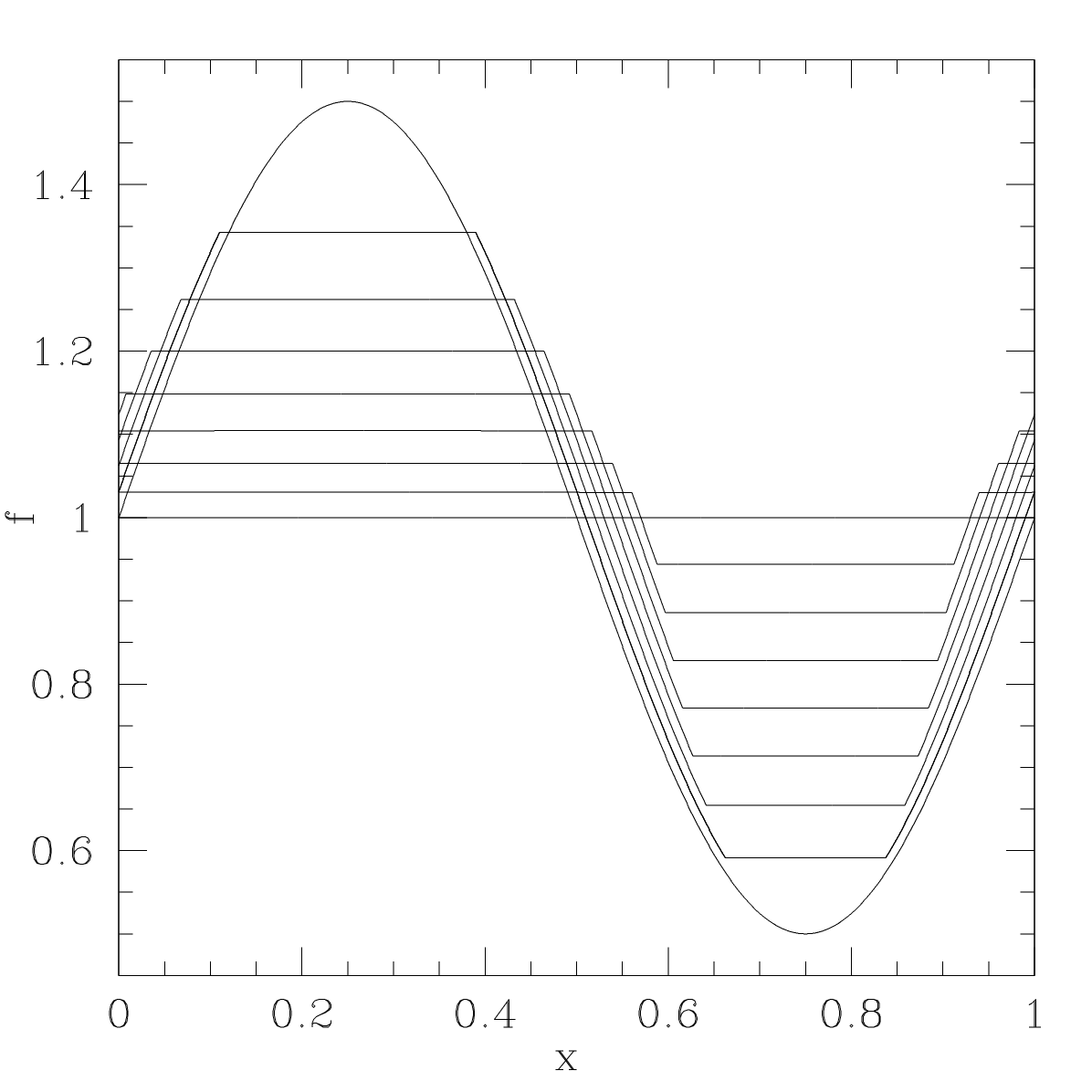}}
\subfigure[]{\psfig{width=2.5in, figure=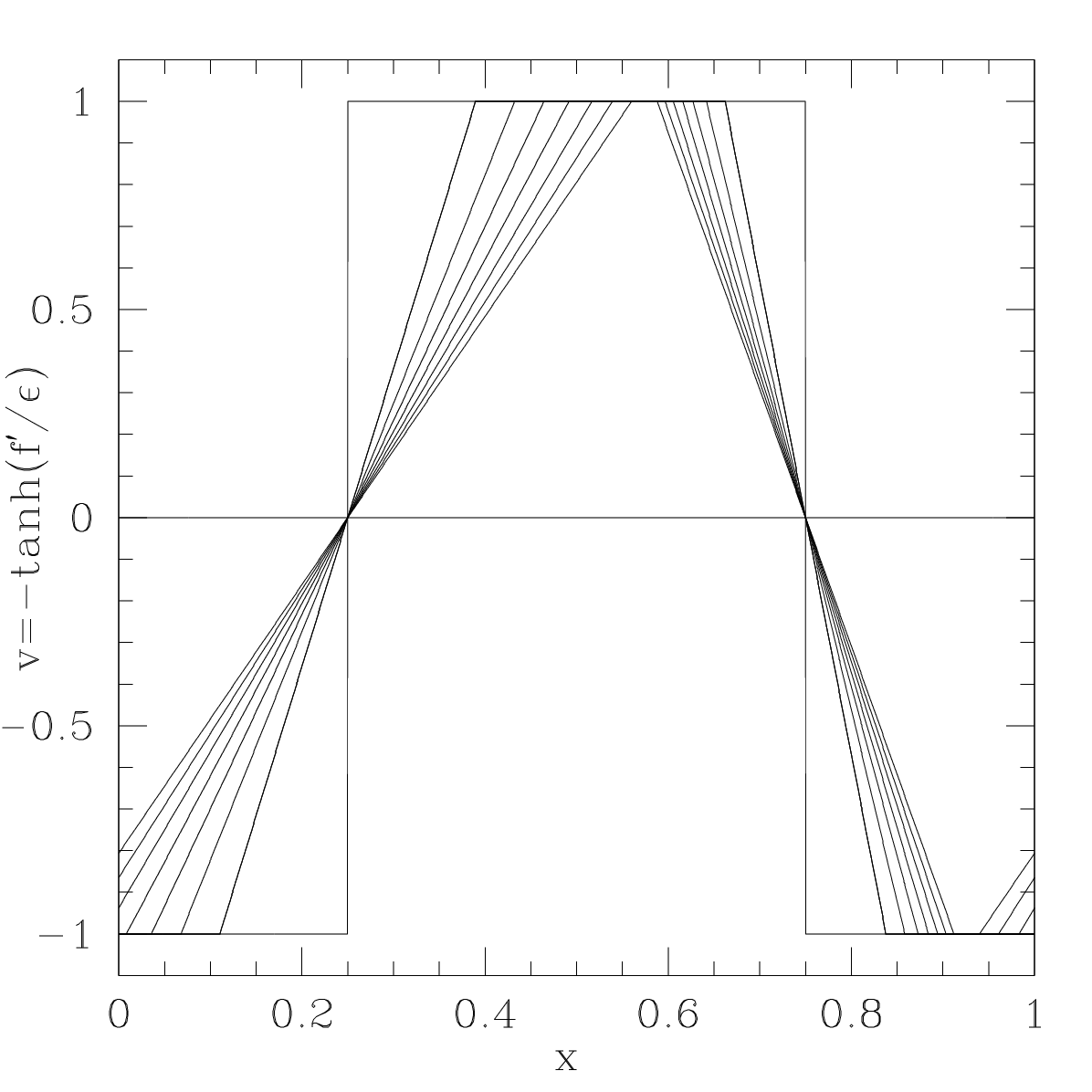}}
\subfigure[]{\psfig{width=2.5in, figure=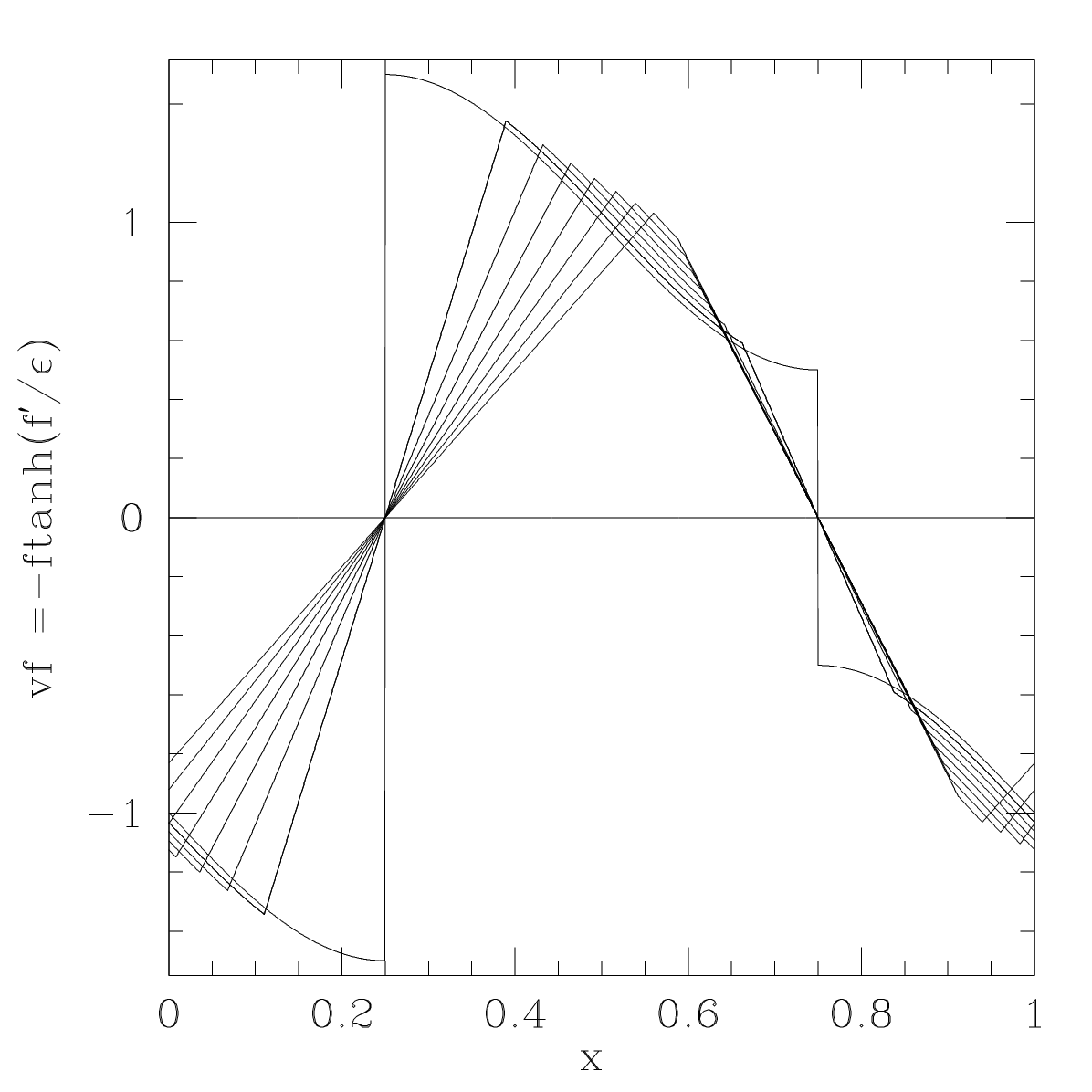}}
\subfigure[]{\psfig{width=2.5in, figure=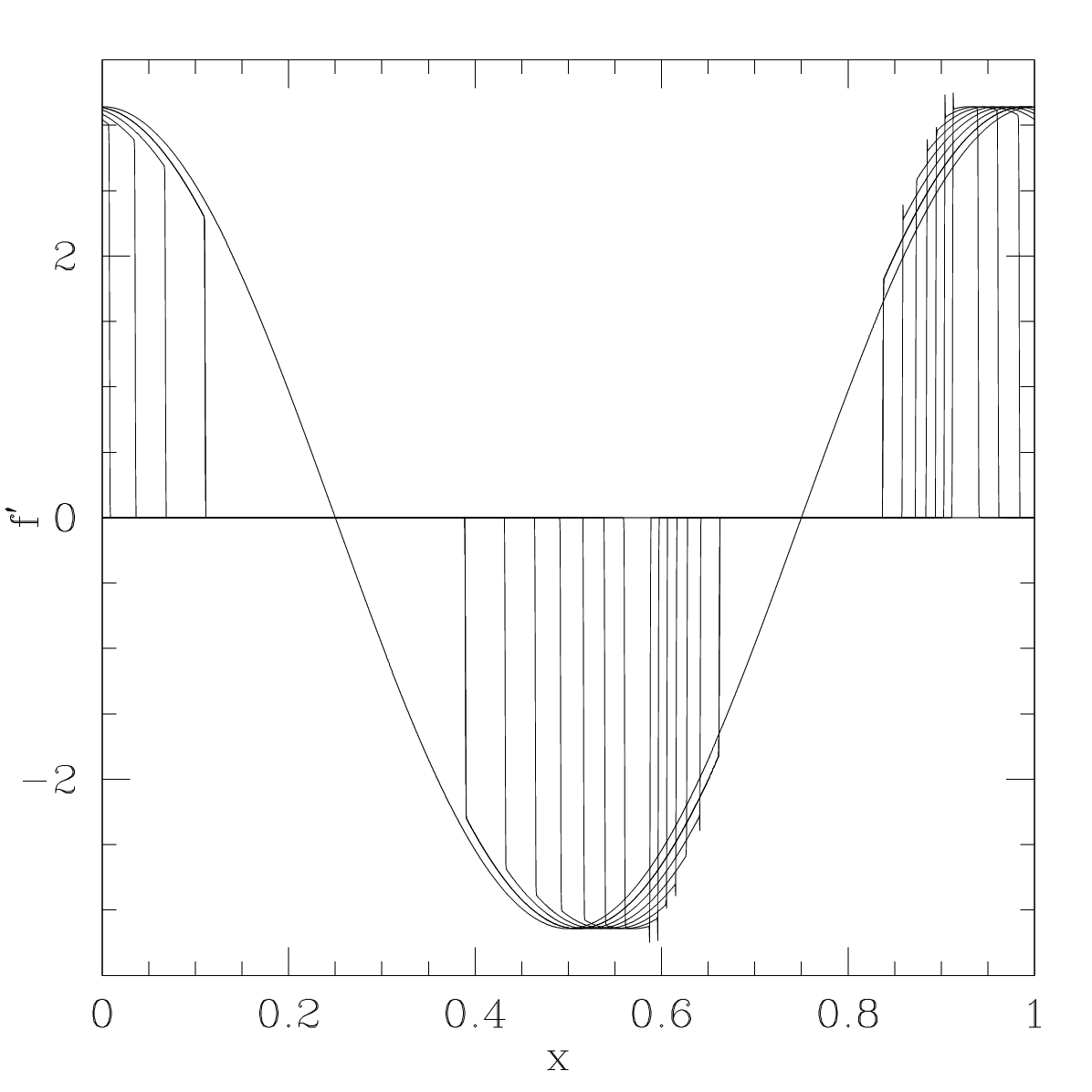}}
\end{center}
\caption{Various quantities from $t=0$ until $t=0.08$ (separated by $\Delta t=0.01$)
as a function of $x$ for the sine wave test problem (Eq. \ref{eq:sin}) using $\epsilon=0.0005$. 
The linearized implicit method with the corresponding stable timestep $\Delta t = 5\times 10^{-7} 
\Delta x$ is used. The number of grid points is 2048. While $f$, $v = -\tanh(f^\prime/\epsilon)$, $vf$ 
are continuous in space, $f^\prime$ is discontinuous where the solution transitions from a flat to a 
steep profile.}
\label{fig}
\end{figure}

\begin{figure}
\begin{center}
\subfigure[]{\psfig{width=2.5in, figure=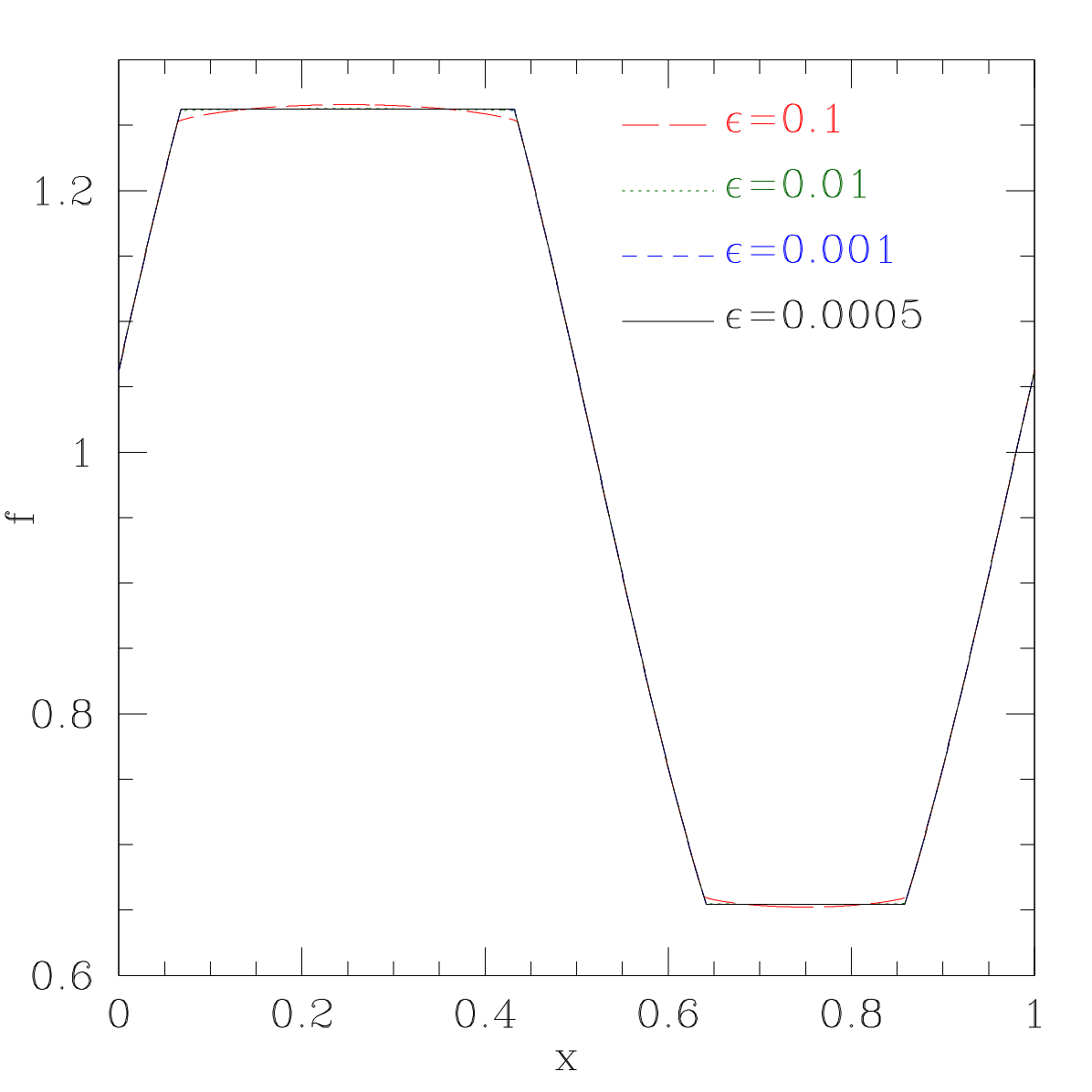}}
\subfigure[]{\psfig{width=2.5in, figure=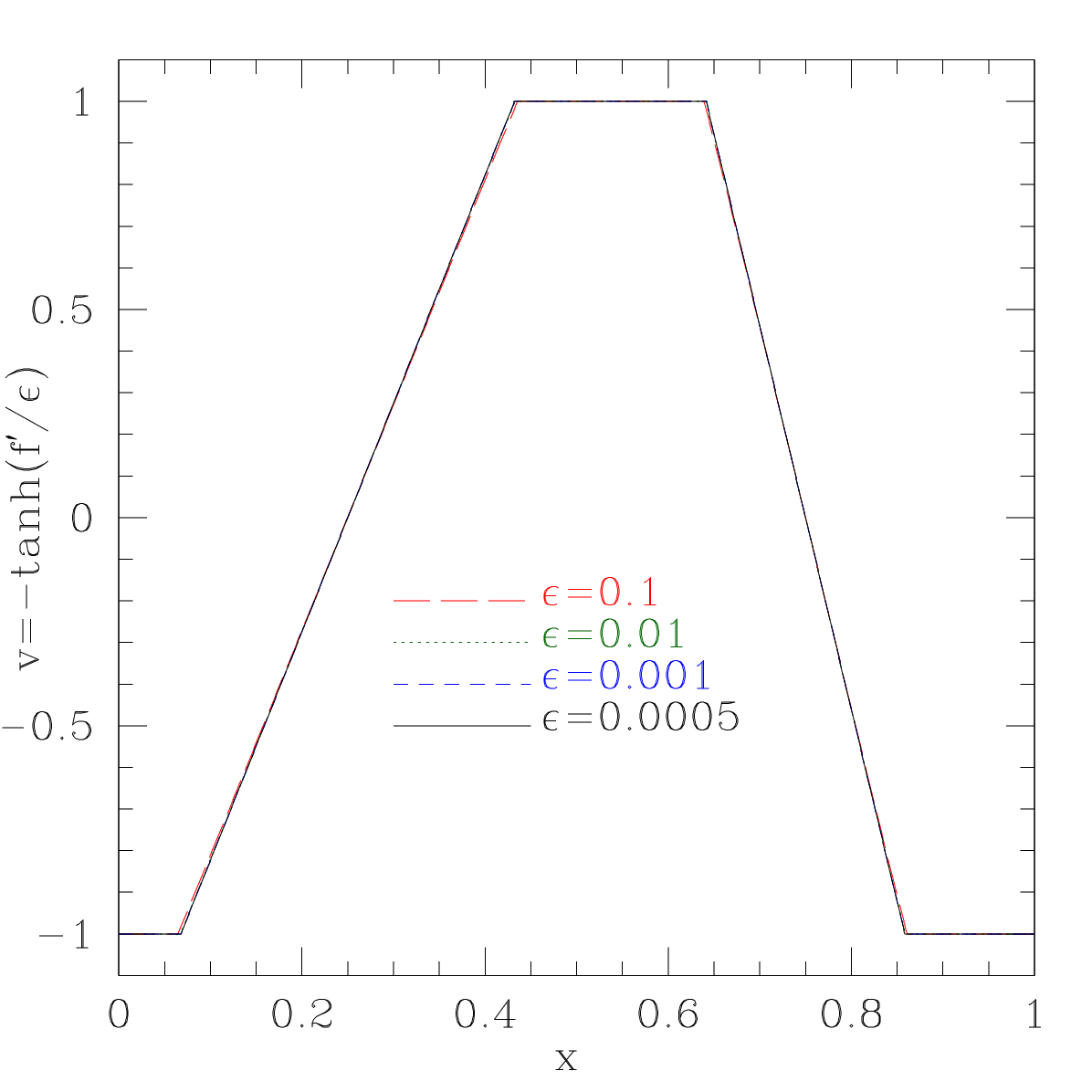}}
\subfigure[]{\psfig{width=2.5in, figure=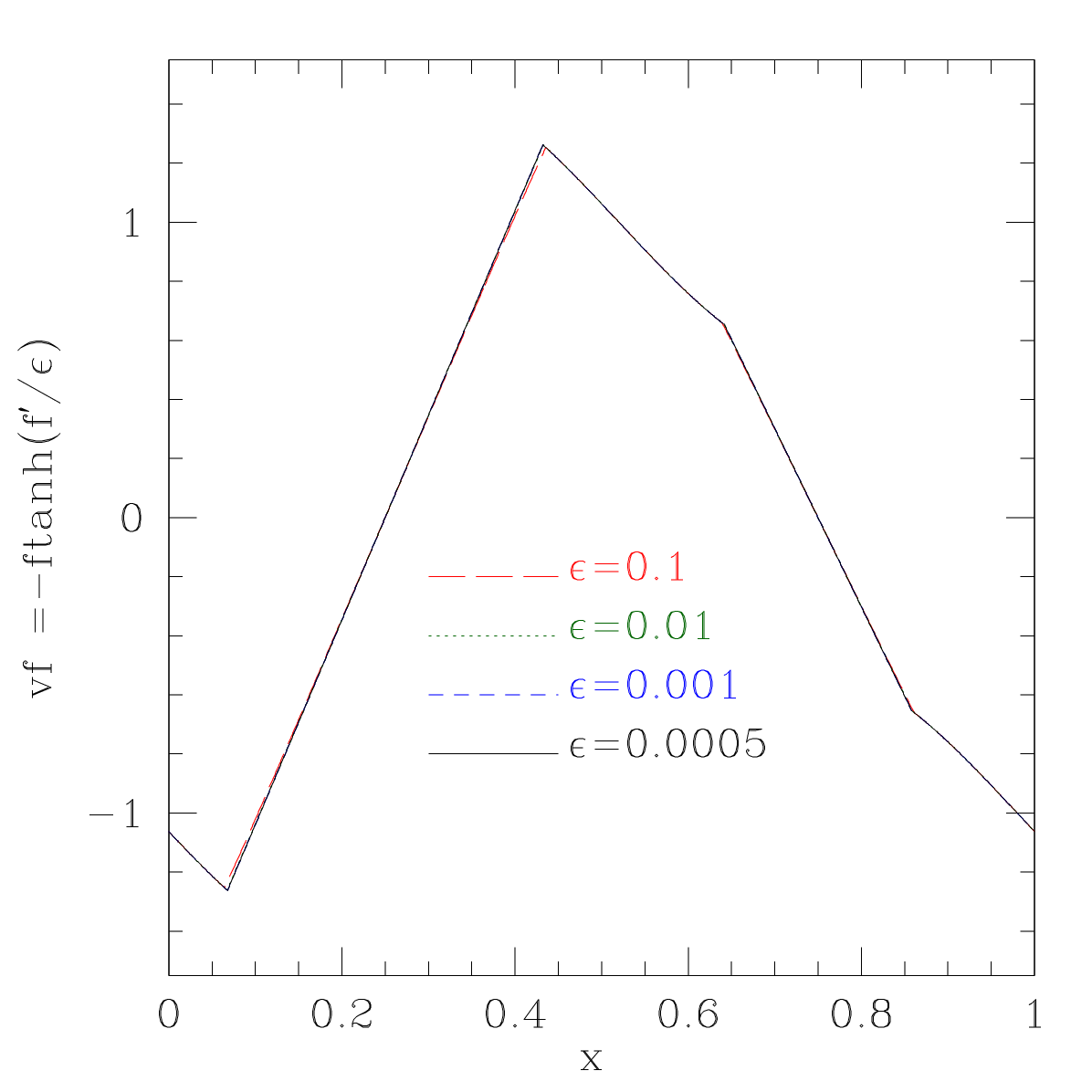}}
\subfigure[]{\psfig{width=2.5in, figure=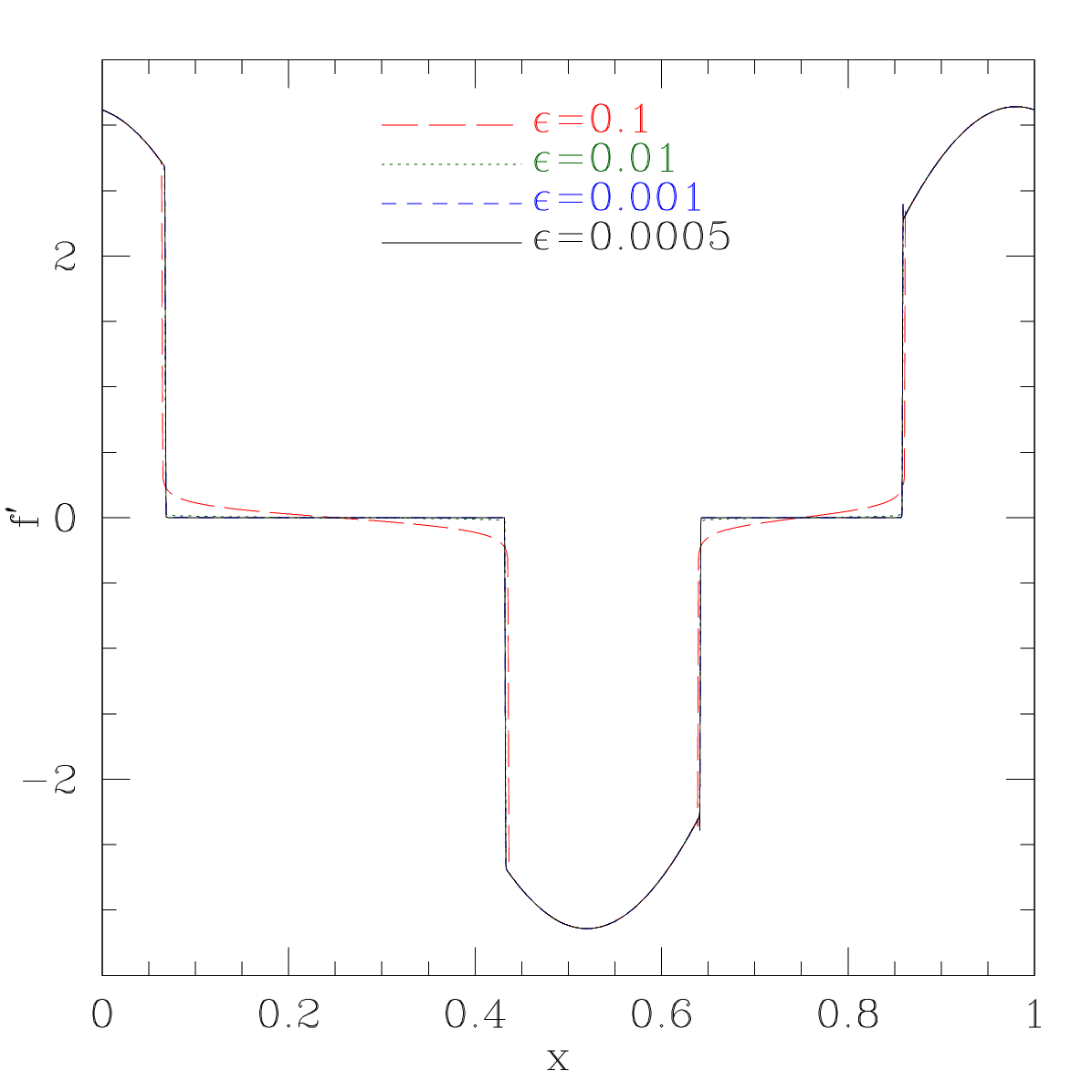}}
\end{center}
\caption{
Various quantities (at $t=0.02$) as a function of $x$ for the sine wave test problem (Eq. \ref{eq:sin}) using $\epsilon=$ 0.1 (blue dashed line), 0.01 (red dotted line), 0.001 (green short-dashed line), \& 0.0005 (solid line). The profiles with different $\epsilon$s lie almost on top of each other.
The linearized implicit method with the corresponding stable timestep ($\Delta t = 5\times 10^{-3}$, $5\times 10^{-5}$, $5\times 10^{-7}$, $5\times 10^{-7} \Delta x$ for $\epsilon=$ 0.1, 0.01, 0.001, $5\times 10^{-4}$, respectively; identical results are obtained for converged solutions with other methods)  is used for different $\epsilon$s. The number of grid points is 2048. Converged profiles are obtained as $\epsilon \rightarrow 0$.}
\label{fig1}
\end{figure}

\subsubsection{Nature of the converged solution}
Now that we have established that quite small timesteps are required to obtain converged results, even for implicit methods, we will examine the nature
of the solutions in more detail. Figure (\ref{fig}) shows the time evolution of various quantities
($f$, $v = -\tanh(f^\prime/\epsilon)$, $vf$, and $f^\prime$) for $\epsilon = 0.0005$ using
the linearized implicit method.
Figure (\ref{fig}a) shows the converged profile of $f$ at different times for the sine wave test problem (Eq. \ref{eq:sin}), using the linearized implicit method. The extrema
of the sine wave are flattened quickly and the $f^\prime \approx 0$ region spreads out in time.  The $f^\prime \ne 0$ portion is advected down the gradient at the streaming velocity, as expected. The initial sine wave is fully diffused out in a short time ($t\approx0.08$). The derivative of the solution changes discontinuously from the flat to steep portions (see Fig. \ref{fig}d).

Fig. (\ref{fig1}) shows the profiles of various quantities ($f$, $v=-\tanh(f^\prime/\epsilon)$, $vf$, and $f^\prime$) at $t=0.02$ for different $\epsilon$s using the linearized implicit method. Timesteps are chosen to be short enough that the solutions are converged for every $\epsilon$. The number of grid point is fixed to be 2048. The profiles of different quantities converge as $\epsilon \rightarrow 0$, as required if the regularized equation (Eq. \ref{eq:reg1}) is well-posed. Although $f^\prime$ changes discontinuously for the converged profile (in regions where solution transitions from being flat to steep), somewhat surprisingly $v=-\tanh(f^\prime/\epsilon)$ is continuous. As expected, $v=1$ and is down the gradient, in the steep portions of the
solution. In flat portions, $v$ smoothly changes from -1 to 1, crossing $0$ where $f^\prime=0$. Although $f^\prime$ appears to be zero throughout the flat portion of the solution, it is strictly zero only at two points corresponding to the extrema. In flat portion of the solution $f^\prime$ is small (but not 0!) and scales with $\epsilon$. The flux of $f$ due to streaming, $vf$, is continuous, and is zero only at points where $f^\prime=0$. 

Numerical solution of the regularized equation (Eq. \ref{eq:reg1}) is analogous to the numerical solution of the Euler/Burger's equations with explicit viscosity. For these equations the discontinuous profile at the shock is resolved by viscous spreading over a finite scale due to explicit viscosity. Thus, the profiles for various fluid quantities are continuous. Similarly, because of regularized diffusion applied where $f^\prime \approx 0$, various key quantities ($f$, $v$, $vf$ shown in Figs. \ref{fig} and \ref{fig1}) become continuous. Similar smoothing of the regularized solution also occurs for an initially discontinuous profile, e.g., the square pulse initial condition discussed in the next section.
Fig. (\ref{fig:fig6}) shows that the solution converges as the resolution is increased for a fixed $\epsilon$. Similarly, Fig. (\ref{fig1}) shows convergence for a fixed number of grid points as $\epsilon \rightarrow 0$. Thus, the solutions of the regularized equation (Eq. \ref{eq:reg1}) converge to a unique solution.

\subsection{Initial square pulse}
In addition to the smooth initial profile, we test different methods with an initial square pulse. Discontinuous cosmic ray pressure profiles can arise
in many circumstances, e.g., supernova shocks. The initial profile is given by
\be
f(x,0) =
\cases {
 2 & {\rm for}$~0.4<x<0.6$,
\label{eq:sqr} \cr
 1  & {\rm otherwise.}  
}
\ee
\begin{figure}
\begin{center}
\subfigure[explicit methods]{\includegraphics[width=2.5in,height=2.5in]{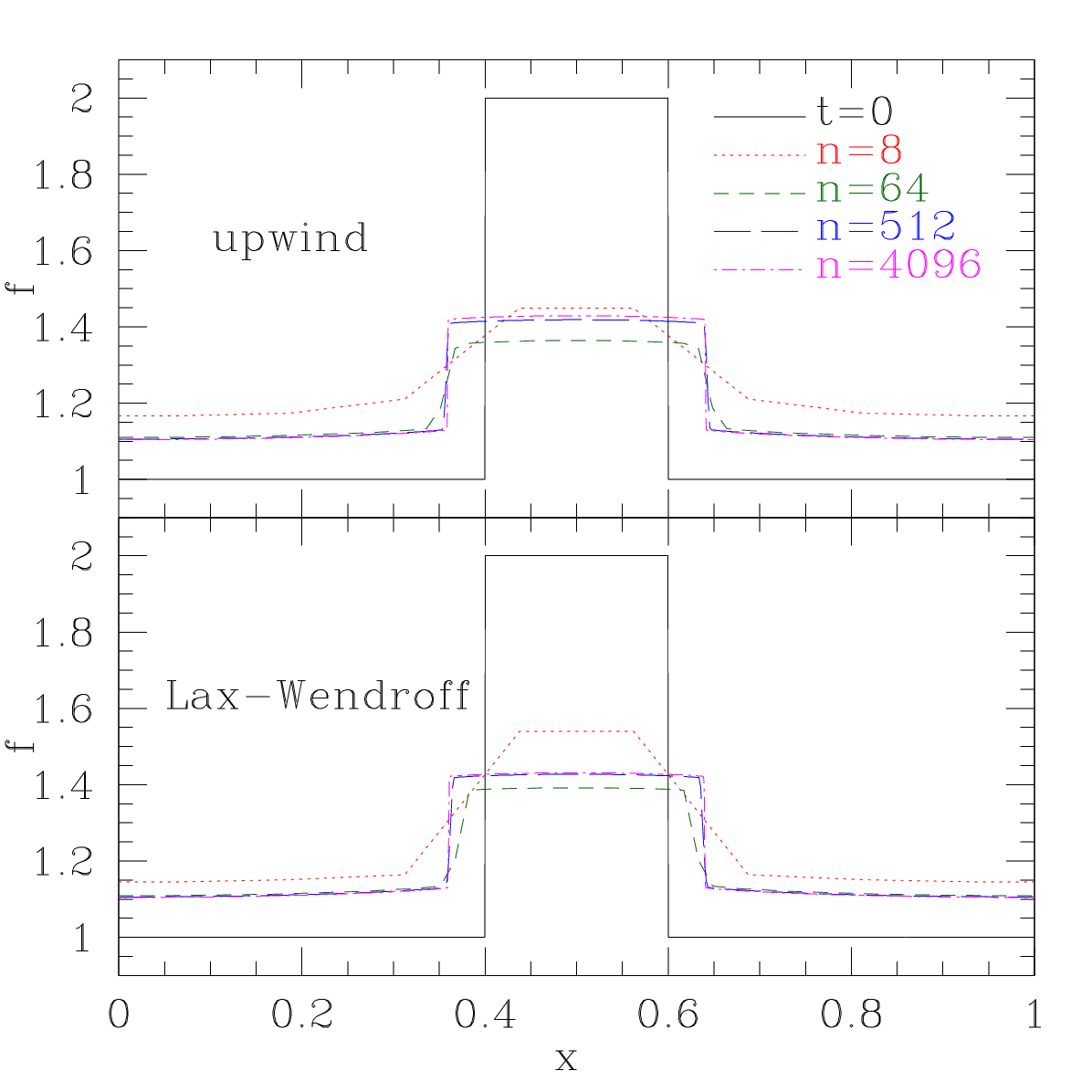}}
\subfigure[implicit methods]{\includegraphics[width=2.5in,height=2.5in]{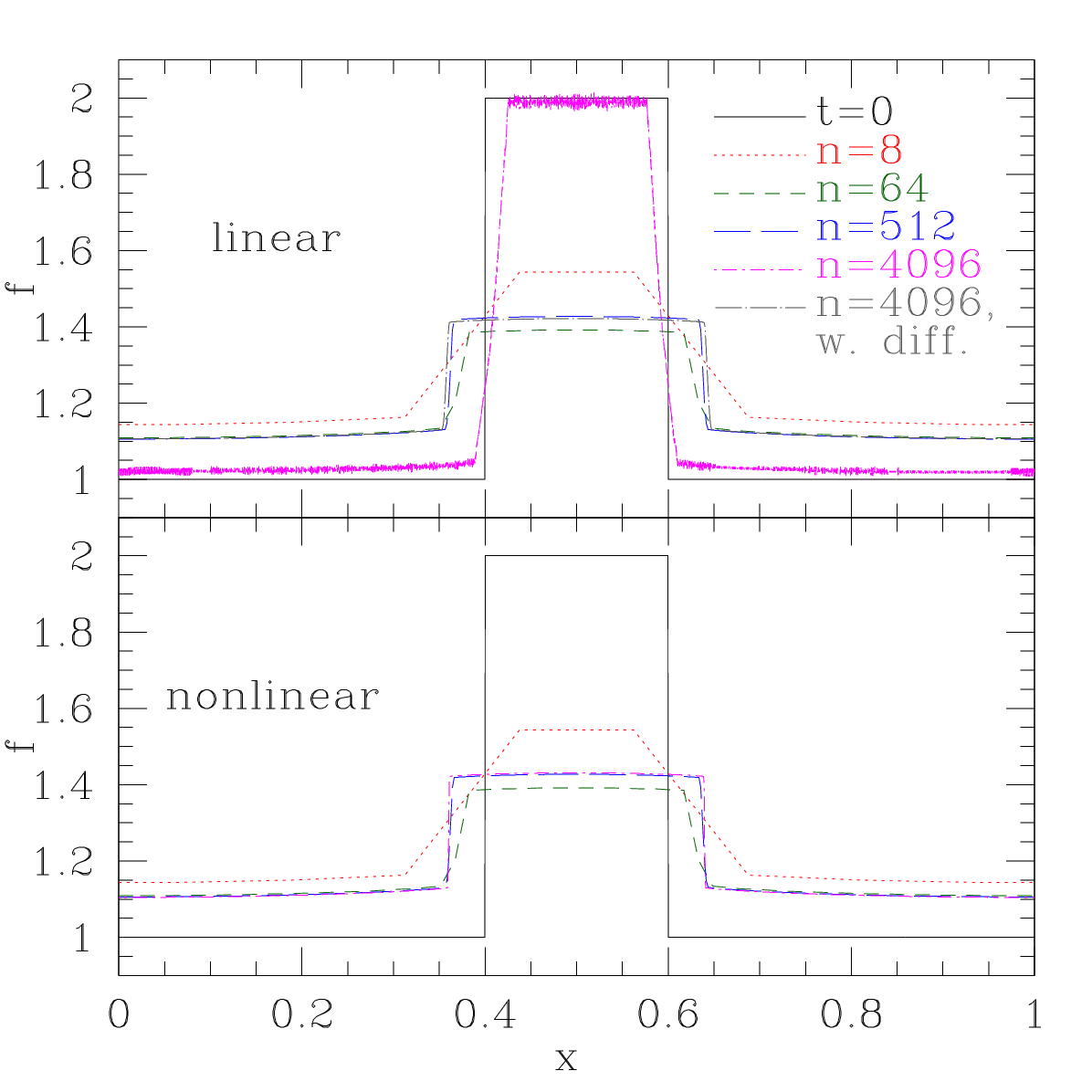}}
\end{center}
\caption{The initial profile (Eq. [\ref{eq:sqr}]) and profiles at
  $t=0.04$ for (a) explicit  and (b) implicit methods at different resolutions with $\epsilon=0.1$. The linearized
implicit method gives unphysical results for higher resolutions. However, converged results are obtained if explicit diffusion (with the diffusion coefficient 
$= \Delta x$) is applied in addition, for the linearized implicit method.}
\label{fig:fig7}
\end{figure}

Figure (\ref{fig:fig7}) shows the solution of Eq. (\ref{eq:reg1}) using explicit (with $\Delta t = \epsilon \Delta x^2/4$, consistent with Eq. [\ref{eq:stab}]; $\epsilon=0.1$) and implicit 
(with $\Delta t= 0.005 \Delta x$) methods discussed in section (\ref{sec:reg}) for the square pulse problem (Eq. [\ref{eq:sqr}]). All methods except
the linearized implicit method give converging results with increasing resolution. Unlike with the smooth initial profile, the linearized method fails 
because it involves taking the derivative of the discontinuous square wave; the linear expansion about $f^\prime$ used in Eq. (\ref{eq:lin_exp}) is not 
valid when $f$ is discontinuous. This spurious behavior for higher resolutions is eliminated if explicit diffusion (with diffusion coefficient $= \Delta x$; recall that 
the maximum streaming velocity equals unity) is 
applied. Thus, the linearized implicit method (or a semi-implicit version of it) which can be solved exactly using a fast tridiagonal solver, is a competitive 
approach even with discontinuous profiles if small explicit diffusion ($\propto \Delta x$) is applied.

\section{Alternate regularizations}
\label{sec:dif_reg}

In addition to using the tanh function to approximate the discontinuous sgn function, we also tried two other functions:  the error function,
\be
{\rm erf}(f^\prime/\epsilon) = \frac{2}{\sqrt{\pi}} \int_0^{f^\prime/\epsilon} e^{-t^2}dt,
\ee
 and a super-exponential function,
\be
{\rm Se}(f^\prime/\epsilon) = 2 e^{-e^{-f^\prime/\epsilon}} - 1.
\ee
While erf is antisymmetric like the sgn function, Se is not.

\begin{figure}
\begin{center}
\subfigure[]{\psfig{width=2.5in, figure=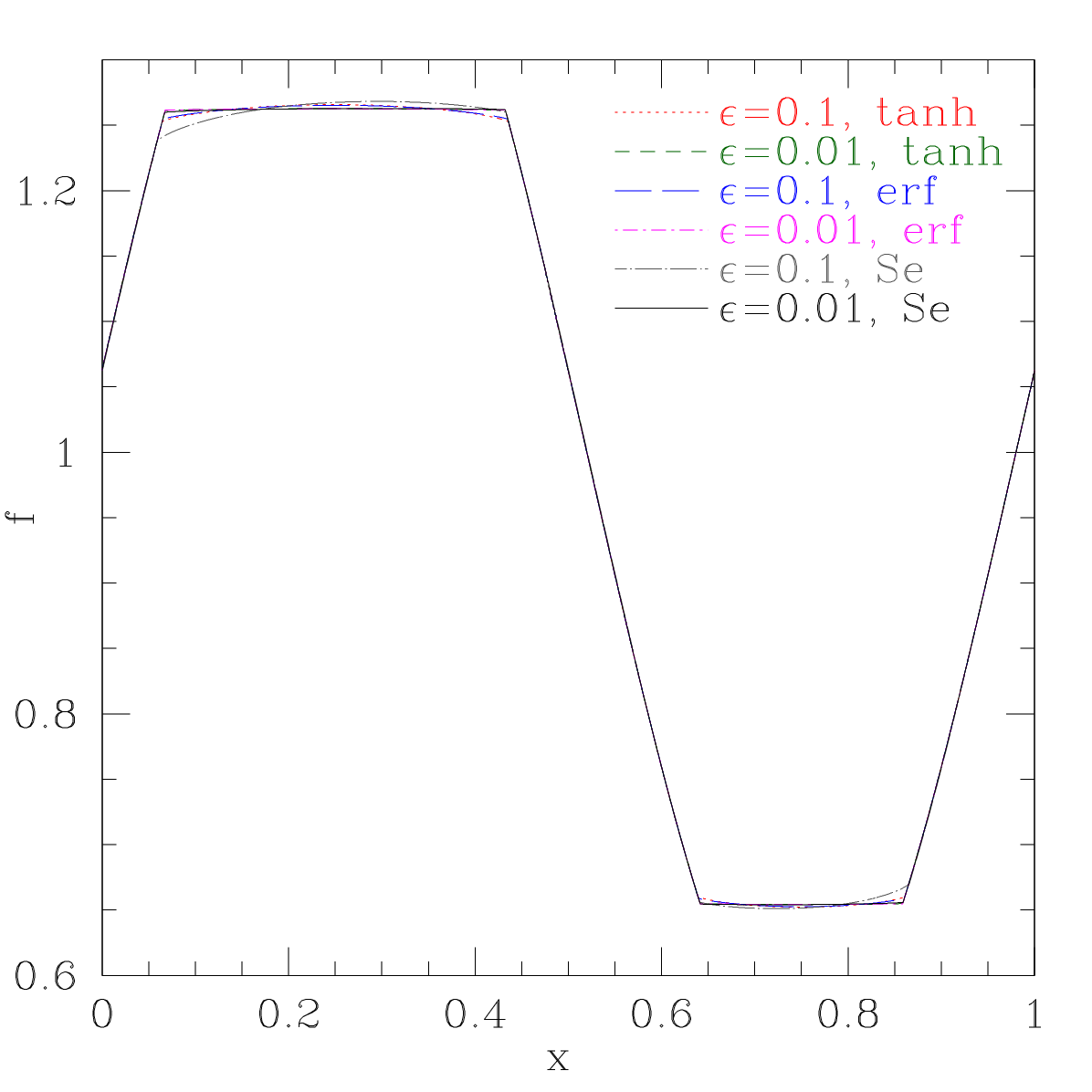}}
\subfigure[]{\psfig{width=2.5in, figure=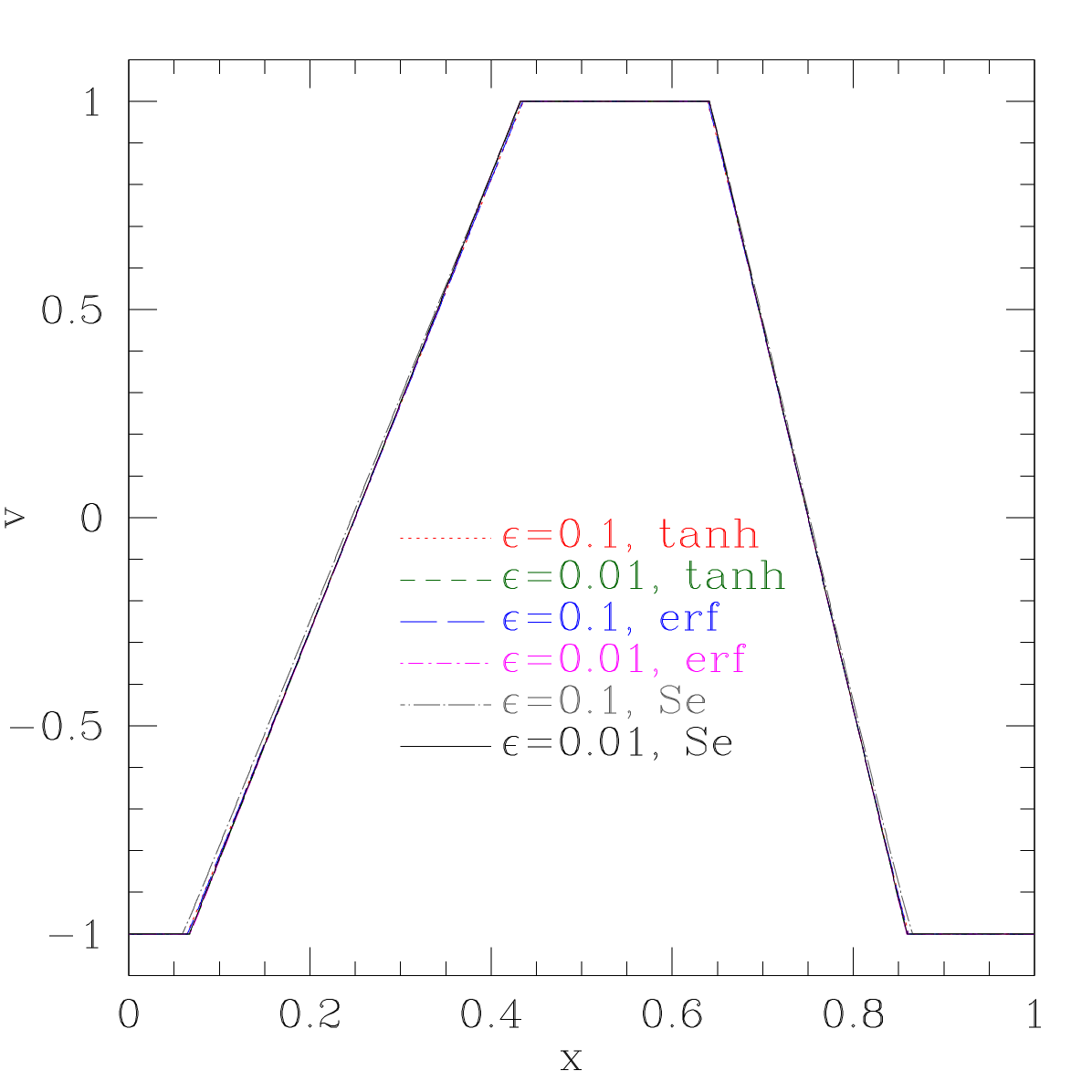}}
\subfigure[]{\psfig{width=2.5in, figure=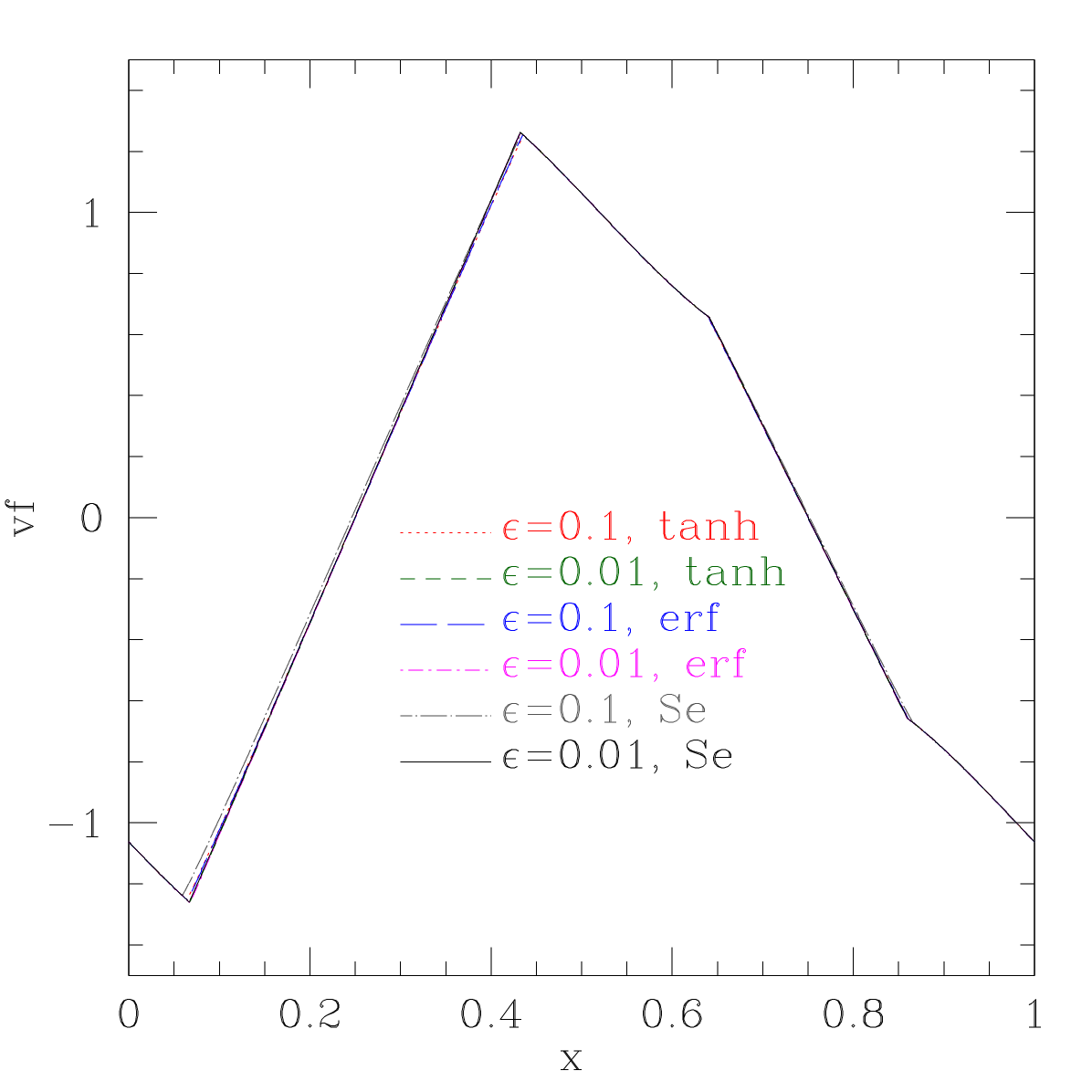}}
\subfigure[]{\psfig{width=2.5in, figure=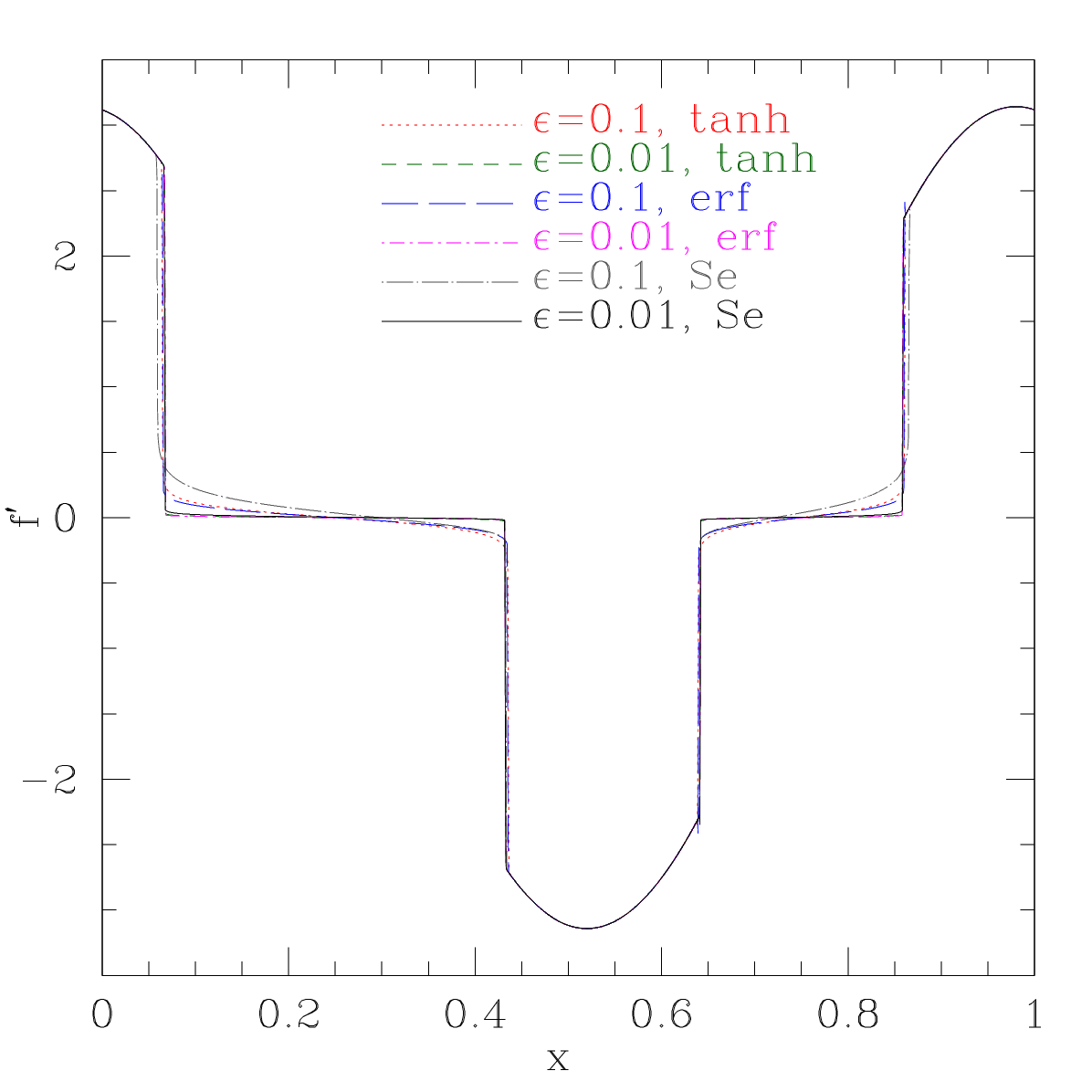}}
\end{center}
\caption{
Various quantities (at $t=0.02$) as a function of $x$ for the sine wave test problem (Eq. \ref{eq:sin}) using $\epsilon=$ 0.1 and 0.01, and with different 
regularizations. Different quantities become closer to each other as $\epsilon$ is reduced. The number of grid points is 2048.}
\label{fig2}
\end{figure}

Figure (\ref{fig2}) compares different quantities at $t=0.02$ using different regularizations and $\epsilon$s. The linearized implicit method is used to evolve
the regularized streaming equation. Timesteps are chosen such that a converged
solution is obtained: for tanh it corresponds to $\Delta t=5\times10^{-3}$, $5\times 10^{-5}\Delta x$ with $\epsilon=0.1,~0.01$ respectively; for erf, $\Delta t=2.5 \times 10^{-3}$, $10^{-5} \Delta x$ with  $\epsilon=0.1,~0.01$ respectively; and for Se, $\Delta t=10^{-3}$, $10^{-5} \Delta x$ with  $\epsilon=0.1,~0.01$ respectively. Thus, the stable timestep depends on the regularization that is used. Notice that for $\epsilon=0.1$ the non-antisymmetric regularization with 
Se results in a profile with is not symmetric with respect to the extrema (see Fig. \ref{fig2}a), but the asymmetry is much reduced for $\epsilon=0.01$.  For 
the regularizations we have tried the result is independent of the regularization in the limit $\epsilon \rightarrow 0$, as is desired.

\section{Conclusions}
\label{sec:conc}
\begin{figure}
\begin{center}
\subfigure[$L=1$, $n=512$]{\includegraphics[width=2.5in,height=2.5in]{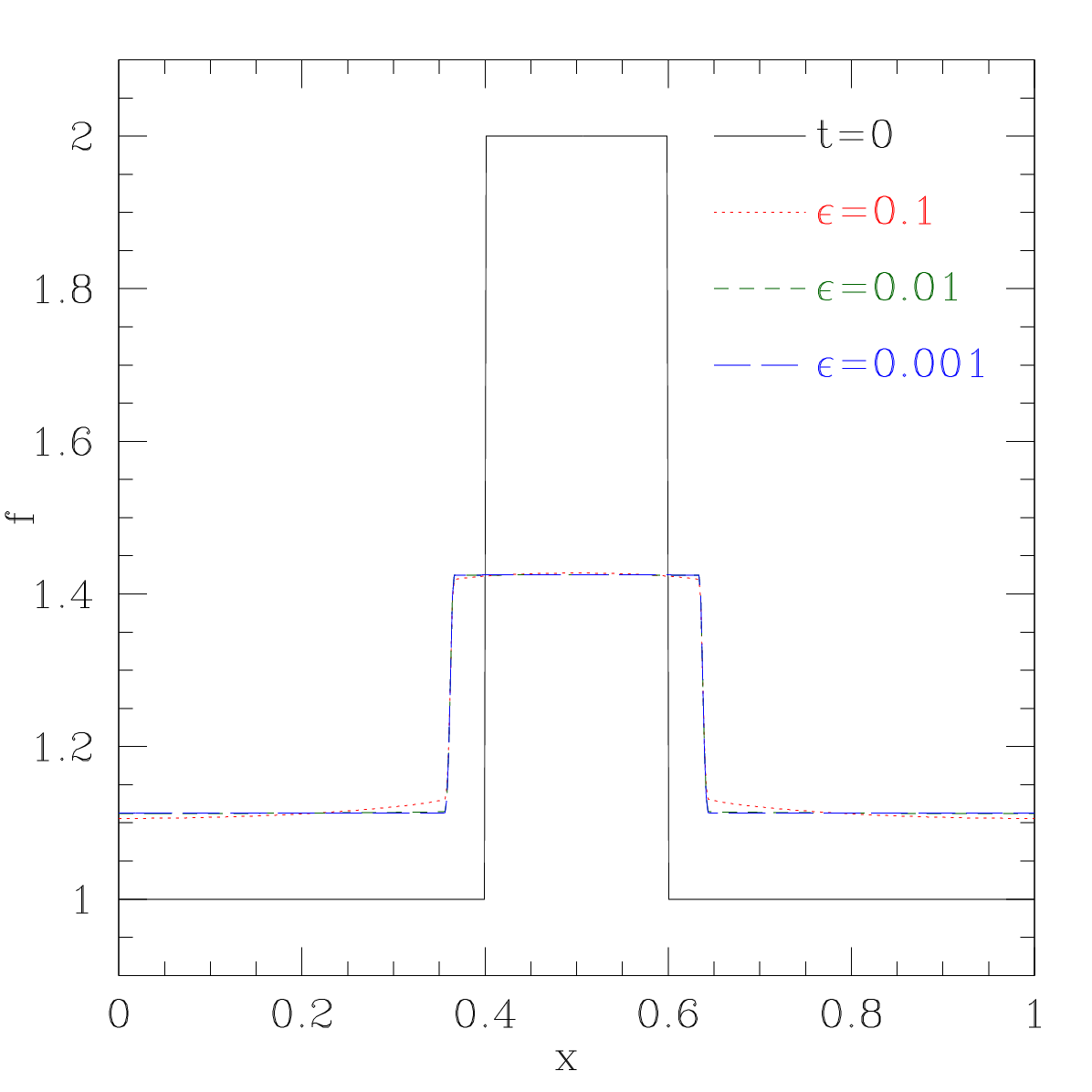}}
\subfigure[$L=16$ and $n=8192$]{\includegraphics[width=2.5in,height=2.5in]{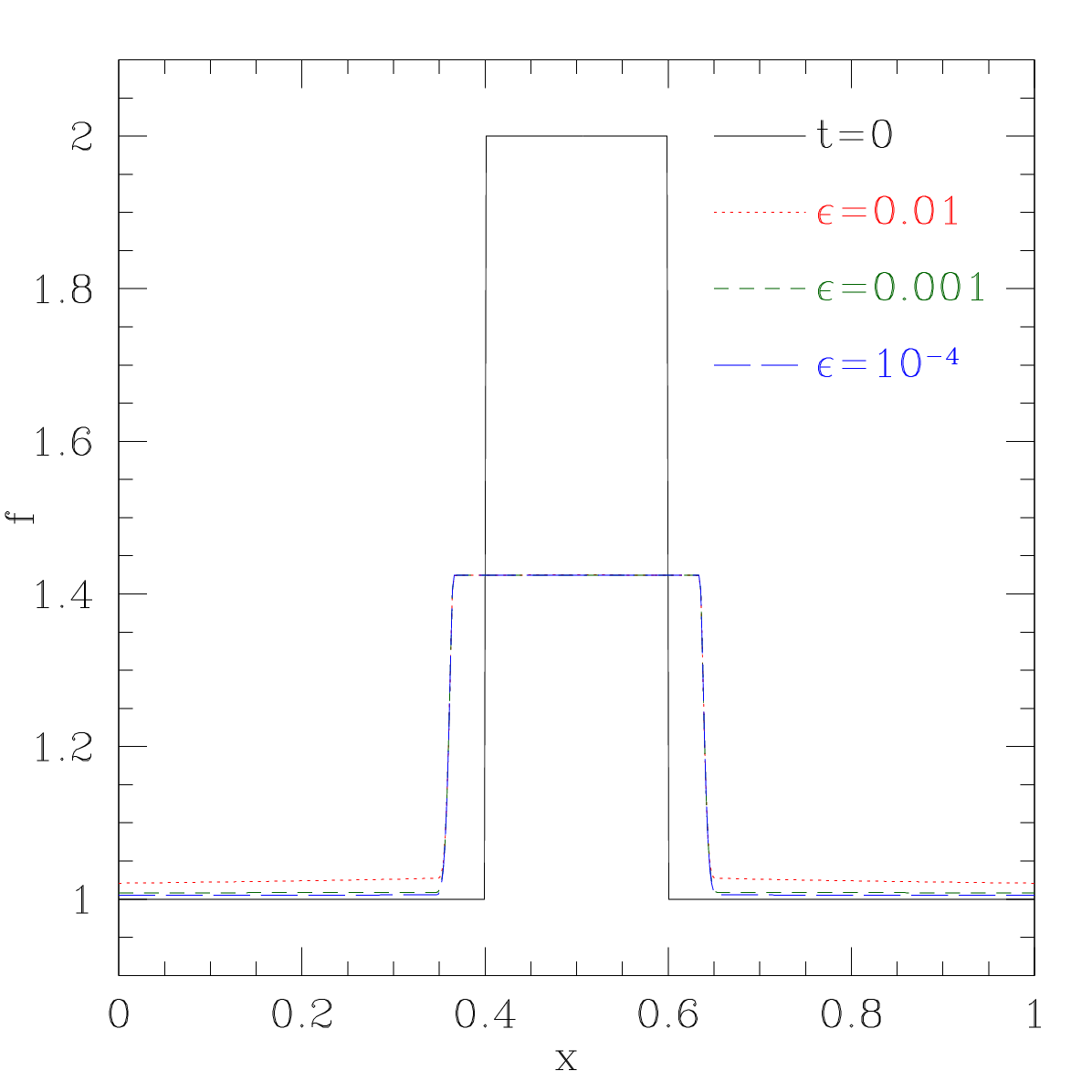}}
\subfigure[$L=32$ and $n=16384$]{\includegraphics[width=2.5in,height=2.5in]{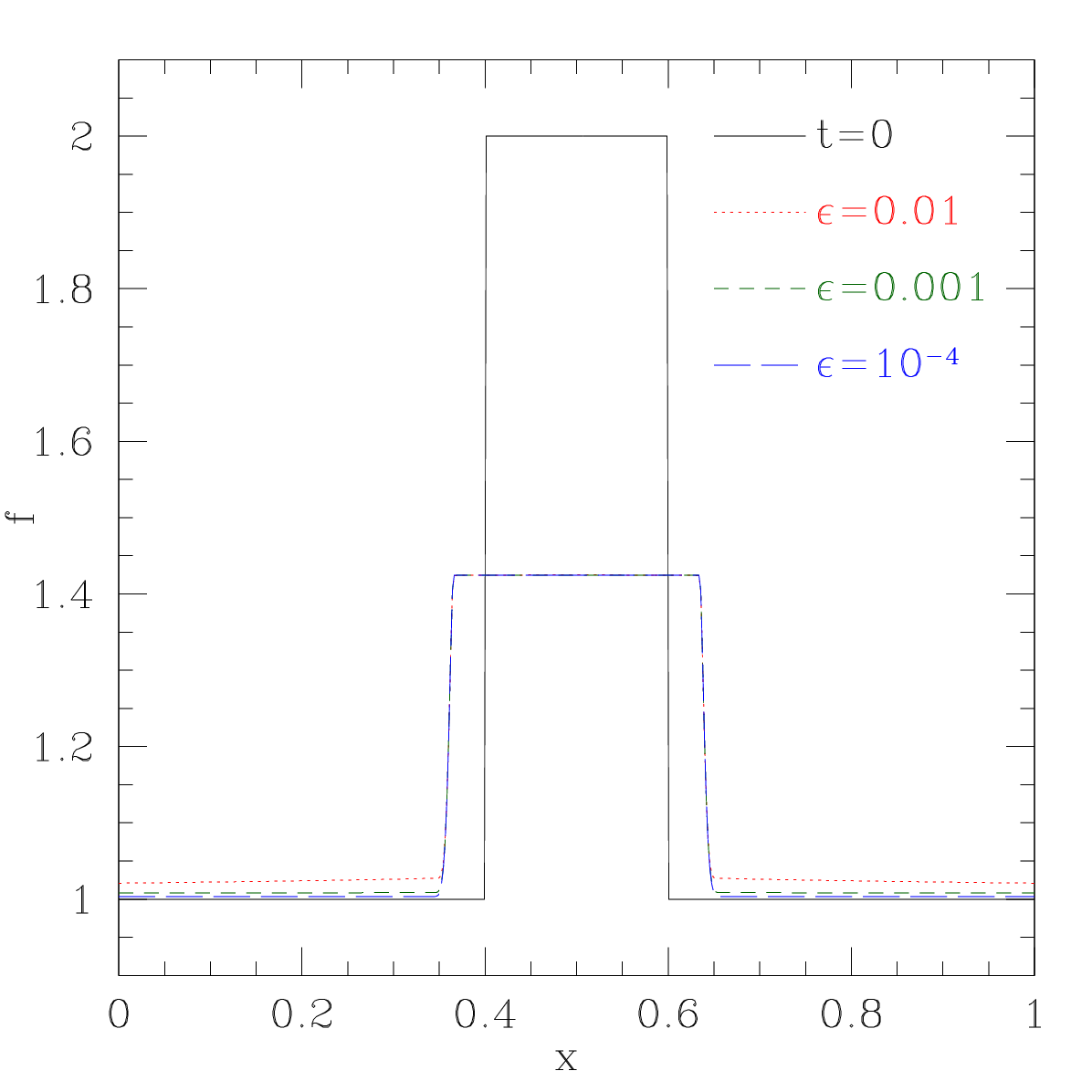} }
\subfigure[$f-1$ {\it vs.} $x$]{\includegraphics[width=2.5in,height=2.5in]{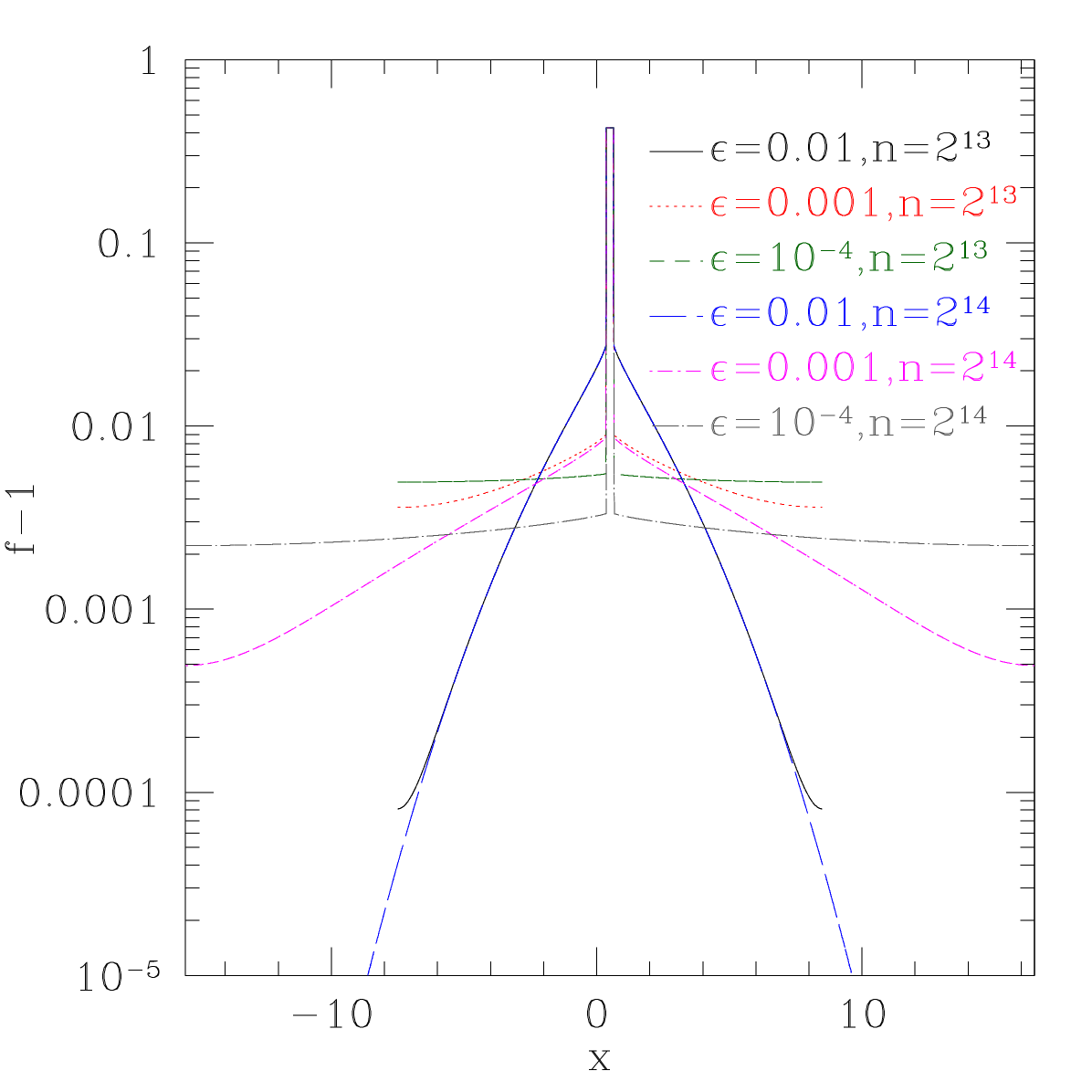}}
\end{center}
\caption{Profiles at $t=0.04$ for the periodic square wave test problem (Eq. [\ref{eq:sqr}]; evolved with the Lax-Wendroff method with $\Delta t = \epsilon \Delta x^2/4$) for different box sizes ($L$), number of grid points ($n$; such that
the number of grid points per unit length is the same for all cases) and  
$\epsilon$s:(a) the profiles with $L=1$, $n=512$ are similar for
$\epsilon < 0.1$; (b) profiles for $L=16$ and $n=8192$ ($2^{13}$); 
(c) profiles for $L=32$ and $n=16384$ ($2^{14}$); (d) logarithmic plot for $f-1$ as a function of $x$ shows that much smaller 
$\epsilon$ is required for convergence with a larger box-size.}
\label{fig:fig8}
\end{figure}

We show that the equation governing streaming down the gradient, although appearing deceptively similar to the advection equation (Eq. \ref{eq:basic}), 
has an entirely different character.
It behaves like a
diffusion equation at the extrema where $f^\prime \approx 0$ (see Eq. \ref{eq:reg}), and like an advection equation away from extrema.
Flux moves from the initial maximum to the initial minimum in the sine wave test problem (initial maximum is reduced and initial minimum is increased) even 
after a short time (see Fig. [\ref{fig}a]) , while the almost flat maximum and minimum are 
connected by a simple advecting profile. This transport of flux to large distances is a clear sign of the non-hyperbolic nature of Eq. (\ref{eq:basic}). Similar signs of large scale transport are seen in the long tails for the square wave problem in Figs. (\ref{fig:fig7}) and (\ref{fig:fig8}). 

The timestep limit for explicit methods on Eq. (\ref{eq:basic}) is severe ($\Delta t \propto \Delta x^3$; see Eq. [\ref{eq:exp_mon}]). However, the regularized equation 
(with a fixed $\epsilon$) is diffusive and the timestep limit for explicit methods scales as $\Delta t \propto \Delta x^2$ (Eq. [\ref{eq:stab}]). The most practical
are the implicit methods, the timestep limit for which scale as $\Delta x$. Although the linearized implicit method gives grid scale oscillations for discontinuous
profiles, a small explicit diffusion cures this problem (see Fig. [\ref{fig:fig7}b]). A fully implicit method using GMRES does not suffer from such problems and does not require explicit diffusion. 
Somewhat unexpectedly, the timestep limit for getting converged solutions does not scale linearly with $\epsilon$; a smaller timestep is required for converging results because of an extremely large diffusion (scaling with $1/\epsilon$) at extrema. It is non-trivial to obtain the timestep limit for the convergence of  implicit methods (and to establish whether $\Delta t \propto \Delta x$ results in a stable scheme as $\Delta x \rightarrow 0$). Traditional
linear stability analysis methods (e.g., von Neumann stability analysis) are not useful because the streaming equation is a highly nonlinear equation even 
in its most basic form (Eq. [\ref{eq:basic}]). 

Total number of operations with $n$, the number of grid points in each direction, scale as: $n^{2+d}/\epsilon$ for explicit 
methods (where $d$ is number of space dimensions); and as $n^{1+d}/\epsilon$ for implicit methods, assuming that each implicit update scales as 
$n^d$ (only if the matrix is well behaved will Krylov methods converge as $n^d$; however the exact tridiagonal solver will definitely scale as $n^d$).  

Although the analysis of a simple equation like (\ref{eq:basic}) is interesting in its own right, the fluid modeling of cosmic rays involves streaming of 
cosmic rays down their pressure gradient and is analogous to the simple streaming equation. 
The generalization to multidimensions is straightforward,
either in directionally split (directional splitting gives accurate results for implicit treatment of 
anisotropic diffusion; see \cite{newref14}) or unsplit fashion. Unlike with splitting, unsplit methods do not result in a tridiagonal matrix but can be solved by fast Krylov subspace methods. One will need to be careful about calculating $\vec{v}_s$ in the numerical implementation of  (\ref{eq:pcr}) in multidimensions;
interpolation of $(\vec{B} \cdot \vec{\nabla})p_c$ using slope limiters (as with anisotropic conduction, e.g., \cite{Sharma2007}) may be required to preserve monotonicity.

The numerical techniques that we have proposed for solving the streaming equation (Eq. [\ref{eq:basic}]) are 
applicable for different initial (even discontinuous initial profiles can be handled; e.g., Fig. [\ref{fig:fig7}]) and boundary conditions. Figure (\ref{fig:fig8})
shows that $\epsilon$ required to obtain converged results scales inversely with the box size, as expected (since $f^\prime/\epsilon \sim f/\epsilon L$  must 
be $\ll 1$ for $\tanh$ to be a good approximation of the sgn function).
While $\epsilon=0.1$ is fine to get profiles similar to those obtained by smaller $\epsilon$s for $L=1$, $\epsilon<0.001$ is required to obtain  the 
$\epsilon \rightarrow 0$ solution for $L=16$ and 32. Notice that the $\epsilon \rightarrow 0$ solution consists of a square pulse and extended flat tails.
Another important point is that the width of the square pulse and the value of $f$ in the 
square pulse at $t=0.04$ is the same for different box sizes and different $\epsilon$s. This implies that the flux leaving the square pulse because of 
advection and diffusion terms in Eq. (\ref{eq:reg}) is independent of $\epsilon$, box size, and the boundary condition. 
If the pulse was just spreading out at a constant velocity (instead of having the extended tails seen in Fig. [\ref{fig:fig8}]), the value of $f$ at $t = 0.04$ within the square pulse would be $f = 1 + 0.2/(0.2 + 2 \times 0.04) = 1.714$, much larger than the value of Å1.425 seen in the converged solutions with streaming. 
The average rate of decrease for $f$ within the square pulse is almost double for the streaming case with tails as compared to a naive advection solution.
Similar large scale transport of flux is seen in the sine wave test problem in Figure (\ref{fig}a).
The relative proportion of fluxes in the advecting square wave and the long tails
depends on the height of the square pulse; higher pulse height leads to a larger flux in the advecting pulse as compared to the extended tails.
Thus realizing the diffusive nature of Eq. (\ref{eq:basic}) can be astrophysically important,
e.g., to accurately obtain the cosmic ray loss rate from the Galaxy and leakage of cosmic rays upstream of supernova shocks.

Comparison of the fluid model (\ref{eq:pcr}) with particle simulations of cosmic rays is required to test if the cosmic ray streaming model applies for a particular situation (e.g., the tails clearly cannot extend to space scales larger than the light crossing time!) but once the
validity of the streaming model for the cosmic ray fluid is established (probably true for Galactic scales), our inferences are applicable. 
In addition to the implicit diffusive nature of the streaming equation, diffusion of cosmic rays because of scattering by Alfv\'en waves (ignored in this paper) can result in additional spreading of cosmic rays; the cosmic ray diffusion time is typically much longer than the streaming time for typical Galaxy and galaxy-cluster parameters (e.g., \cite{Loewenstein1991}). 
Streaming of cosmic rays at the local Alfv\'en speed is extremely important for cosmic ray confinement in the Galaxy for millions of years. Cosmic
rays  can heat the interstellar 
and the intracluster medium (due to the $|v_s\cdot \vec{\nabla} p_{\rm cr}|$ term on the right hand side of Eq. [\ref{eq:pcr}] or because of 
convection driven by them) where cosmic ray pressure (compared to the thermal plasma pressure) is significant (e.g., \cite{Guo2008,Sharma2009}).  
The often-used equation of saturated thermal conduction, thermal conduction at length scales larger than the electron mean free path, given by (e.g., Eq. 2.8 in \cite{Balsara2008})
\be
\frac{\partial T}{\partial t} = \vec{\nabla} \cdot \left ( \hat{b}f_s c_s T {\rm sgn}[\hat{b}\cdot \vec{\nabla} T] \right ),
\label{eq:sat_con}
\ee
is very similar in nature to the streaming equation (Eq. \ref{eq:basic}). Here $T$ is the plasma temperature, $c_s \propto \sqrt{T}$ is the sound speed, $\hat{b}$ is the magnetic field unit vector, and $f_s$ is a factor of order unity. The presence of sgn function in the heat flux results in a diffusive behavior at extrema. Thus, like the streaming equation, Eq. \ref{eq:sat_con} is also parabolic in nature, unlike what earlier analyses which have overlooked the nature at extrema have suggested (e.g., \cite{Balbus1986}).
Thus, the methods developed in this paper will be useful in modeling various astrophysical plasma effects.

Solving the regularized form (Eq. \ref{eq:reg1}) of the streaming equation (Eq. \ref{eq:basic}) results in a unique, converged solution, provided that the
timestep is small enough. Regularization is analogous to introducing explicit viscosity in solving Euler/Burger's equations and resolving the viscous length scales at the shocks. With regularization the solutions are approximated very accurately, the advection velocity ($v\equiv -\tanh(f^\prime/\epsilon)$) and streaming fluxes ($vf$) are continuous, 
and converged results are obtained in the limit $\epsilon \rightarrow 0$ and $\Delta x \rightarrow 0$ (e.g., see Figs. \ref{fig:fig6} \& \ref{fig1}). In this limit 
the solutions are also independent of different regularizations we tried (see Fig. \ref{fig2} and \S \ref{sec:dif_reg}).

More work is clearly needed to understand the nature (both numerically and analytically) of the streaming equation (Eq. \ref{eq:basic}). For example, like
the Euler equations which can result in discontinuous profiles starting with a smooth initial condition, a smooth initial profile for the streaming equation can give rise to discontinuous first derivatives (e.g., $f^\prime$ changes discontinuously from $\approx 0$ within the flattened sine wave [e.g., Fig. \ref{fig}] to a constant 
value outside of it). There is significant transport over large scales (e.g., tails in Fig. [\ref{fig:fig8}]) because of the parabolic nature of the streaming 
equation at extrema.

\section{Acknowledgements}
PS was supported by NASA through Chandra Postdoctoral Fellowship grant number PF8-90054 awarded by the Chandra X-ray Center, which is 
operated by the Smithsonian Astrophysical Observatory for NASA under contract NAS8-03060. This research was supported in part by the National Science Foundation through TeraGrid resources provided by NCSA and Purdue University. We thank Greg Hammett for providing the GMRES implementation that is used in this paper. PS thanks Eliot Quataert for encouragement, and Xylar Asay-Davis and Ian Parrish for useful discussions. 
Phil Colella and Dan Martin were supported by the US Department of
Energy Office of Advanced Scientific Computing Research under contract number
DE-AC02-05CH11231. We thank the anonymous referees for comments which significantly improved the quality of the paper.

\end{document}